\documentclass[usenatbib]{mnras}

\usepackage[T1]{fontenc}
\usepackage{ae,aecompl}
\usepackage{graphicx}
\usepackage{hyperref}
\usepackage{amsmath}
\usepackage{amssymb}
\usepackage{mathtools}
\usepackage{bm}
\usepackage{cleveref}

\title[The role of mass and superfluid reservoir in pulsar glitches]{The role of mass, equation of state and superfluid reservoir in large pulsar glitches}

\author[A. Montoli, M.Antonelli \& P.Pizzochero]{
    A.~Montoli$^{1,2}$ \thanks{E-mail: alessandro.montoli@unimi.it},
    M.~Antonelli$^{3}$, 
    P.~M.~Pizzochero$^{1,2}$\\ \\
    $^{1}$Dipartimento di Fisica, Universit\`a degli Studi di Milano, Via Celoria 16, 20133 Milano, Italy\\ 
    $^{2}$Istituto Nazionale di Fisica Nucleare, sezione di Milano, Via Celoria 16, 20133 Milano, Italy\\
    $^{3}$Nicolaus Copernicus Astronomical Center of the Polish Academy of Sciences, Bartycka 18, 00-716 Warszawa, Poland}

\begin{document}
    %
    %
    \pagerange{\pageref{firstpage}--\pageref{lastpage}} \pubyear{2019}
    \maketitle
    \label{firstpage}

\begin{abstract}

Observations of pulsar glitches may provide insights on the internal physics of neutron stars and recent studies show how it is in principle possible to constrain pulsar masses with timing observations. 
The reliability of these estimates depend on the current uncertainties about the structure of neutron stars and on our ability to model the dynamics of the superfluid neutrons in the internal layers.  
We assume a simplified model for the rotational dynamics of a neutron star and estimate an upper bound to the mass of 25 pulsars from their largest glitch and average activity: the aim is to understand to which extent the mass constraints are sensitive to the choice of the unknown structural properties of neutron stars, like the extension of the superfluid region and the equation of state.
Reasonable values, within the range measured for neutron star masses, are obtained only if the superfluid domain extends for at least a small region inside the outer core, which is compatible with calculations of the neutron S-wave  pairing gap. 
Moreover, the mass constraints stabilise when the superfluid domain extends to densities over nuclear saturation, irrespective of the equation of state tested. 
\end{abstract}



\section{Introduction}
\label{sec:intro}

In the current description of pulsar glitches - sudden spin-ups observed in the otherwise steadily decreasing rotational frequency - the neutron star is assumed to be divided in two components  that can rotate with slightly different angular velocities \citep{haskellmelatos2015}: a normal component, which rotational period can be tracked by observing the pulsar electromagnetic emission, and a superfluid component (consisting of paired neutrons) filled by a large number of quantised vortices \citep{haskellsedrakian2017, chamel_review2017JApA}. 
The possibility of pinning between vortices and impurities in the inner crust forces the superfluid to lag behind the normal component in its spin-down \citep{andersonitoh1975} and a superfluid current develops in the frame of the crustal lattice. 
Part of the angular momentum associated with this neutron current is then released during a glitch: unpinned by a still unknown trigger mechanism, vortices suddenly transfer the stored angular momentum to the normal component, causing a glitch.

This set of ideas must be confronted to a vast phenomenology: measured glitch sizes span several decades \citep{espinoza+2011}  and among active pulsars some have shown only glitches of approximately the same size, while others do not seem to have a preferred amplitude \citep{melatos2008,howitt2018ApJ}. 
In some cases, a pulsar can show a single large glitch and several other events orders of magnitude smaller, while the glitch size distribution of the whole pulsar population is found to be bimodal \citep{fuentes17}, a fact that may be explained by invoking different types of glitch triggers or mechanism. 

One of the open problems in the two-component scenario is understanding where the superfluid reservoir involved in the glitch is located.
Early on, \cite{baym+1969} proposed the simplest phenomenological model to describe glitches by considering two rigidly rotating components: a plasma of charged particles (nuclei in the crust, protons in the core and electrons), and a neutron superfluid extending from the drip point in the crust to the whole core. 
However, because of the entrainment coupling - first introduced in the context of superfluid helium \citep{Khalatnikov57,andreevbashkin1975} -  between the neutrons and the proton superconductor in the core, the quantized vortex lines permeating the neutron superfluid are magnetized and can interact with the electron fluid \citep{Sedrakian1980}.
Accounting for this additional mechanism, the neutron superfluid in the core is expected to be strongly coupled to the normal component on timescales of the order of the second \citep{alpar84rapid}.
Therefore, only the dripped neutrons permeating the inner crust have been thought to store the angular momentum needed to spin-up the normal component. 
This scenario was also supported by the analysis of the average glitch activity in the Vela pulsar, from which it has been deduced that the moment of inertia of the internal region related to the angular momentum reservoir for glitches is of the order or few percent of the moment of inertia of the whole star \citep{dattaalpar1993, LE99}.

More recently,  it has been shown that only the unbound ``conduction'' neutrons can move freely past through the ionic lattice formed by the nuclei \citep{carter2005NuPhA}.
In hydrodynamic models of the inner crust superfluid, this phenomenon can still be described as an entrainment coupling (of a different kind with respect to the one arising in the core) that changes the bare neutron mass to an effective mass of the ``free'' neutrons \citep{carter2006IJMPD}.
In particular, \cite{chamelcarter2006} derived how the entrainment effect leads to a rescaling of the moments of inertia of the superfluid and normal components. As a consequence, because of the large estimates of the entrainment coupling in the crust \citep{chamel2012} it is impossible to explain the glitch activity of the Vela pulsar if the angular momentum reservoir is confined into the crust \citep{andersson+2012, chamel2013, delsate2016}.

This points to the direction of considering a superfluid reservoir that extends into the core, either by invoking a collective rigidity of the vortex bundle \citep{ruderShut1974,pizzochero2011} or by invoking additional pinning mechanism of the vortex lines to the flux-tubes in the core \citep{gercinog2014}.

To overcome the difficulty posed by strong crustal entrainment, it is also possible that the only type of superfluid involved in the glitch phenomenon is that in the singlet ${}^1S_0$ state, which in several theoretical calculations extends beyond the crust-core boundary. In this case, the superfluid reservoir also depends on the model used for the superfluid gap and on the internal temperature of the neutron star, whose estimate is based on the inferred age of the pulsar and on the particular cooling model used \citep{ho+2015}. 

In any of the cases considered above, the moment of inertia of the pinning region depends on the unknown properties of dense matter near and above nuclear saturation density. In fact, different equations of state (EoSs) imply different structural proprieties of the star such as maximum mass, crustal thickness and free neutron fractions. 
These properties have been implemented in a simplified hydrodynamical model to set an upper limit on the mass of a glitcher by means of its largest observed glitch amplitude $\Delta \Omega$ \citep[][hereafter Paper-I]{pizzochero+2017}. 

In addition to the calculation of the mass upper bound, $M_{\rm max}$, in Paper-I also a refined upper bound, $M_{\rm act}$, has been proposed, by employing its largest observed glitch and a mean waiting time between large glitches determined from the pulsar's activity. In Paper-I, the estimates have been made in the particular case of a superfluid reservoir extended to the whole star. 
The aim of the present work is to relax this hypothesis, by evaluating the dependence of the mass estimates on different extensions of the S-wave superfluid domain, thus implicitly considering the possibility of different superfluid gaps for the ${}^1S_0$ state, similarly to what has been discussed by \citet{ho+2015}. 

First, we identify a sample of glitchers with observational criteria that allow us to determine the typical timescale between two large glitches. Then, we apply our model to the pulsars in the sample and study how the distribution of the mass estimates for these objects varies for different extensions of the $S$-wave gap. Finally, we explore the effect of different equations of state on the mass estimates.


\section{Mass upper bounds: general approach}
\label{sec:model}

In the following we present the general approach underlying the specific toy-model presented in Paper-I. 
The basic idea is to follow the evolution of the angular momentum reservoir during the spin-down phase in a pulsar. 
This allows to obtain a theoretical time-dependent upper bound on the observed glitch amplitude, which has to be compared with the observed timing properties of a given pulsar. 

\subsection{Evolution of the maximal glitch amplitude}
\label{susec:following}

To uniform with previous works, we  indicate with the subscript $p$ the quantities related to the normal component (the crustal lattice and everything tightly coupled to it), which is assumed to be rigid \citep{Easson1979}; the subscript $n$ is used to indicate the superfluid neutrons, which are treated as a fluid component and can develop non-uniform rotation  \citep[e.g. ][]{prix2002,andersson2006CQGra,sidery2010MNRA,hasekll2012}.

In the absence of precession, the total angular momentum $L$ of a slowly rotating neutron star in General Relativity can be split as \citep{antonelli+2018}
\begin{equation}
L = I\Omega_p + \Delta L [\Omega_{np}] \, ,
\label{eq:ang_momentum}
\end{equation}
where $I$ is the relativistic total moment of inertia of the star \citep[in the sense provided by][]{hartle67}. 
The angular velocity of the rigidly rotating normal component, as seen from an inertial observer at spatial infinity, is $\Omega_p$. 
In the above expression, $\Delta L$ represents the extra angular momentum due to the presence of a non-uniform velocity lag $\Omega_{np} = \Omega_n - \Omega_p$ between the two components. 
Moreover, in the slow rotation approximation $\Delta L$ is a linear functional of the lag $\Omega_{np}$. 
Neglecting a possible time dependence of the metric (in particular of the relativistic frame drag) and of the moment of inertia $I$ (namely, ruling out the possible occurrence of structural changes), we have that 
\begin{equation}
I\dot{\Omega}_p + \Delta L [ \, \partial_t \, {\Omega}_{np} \, ] = - I |\dot{\Omega}_\infty| \, ,
\label{eq:cons_ang_momentum}
\end{equation}
where we can bring the partial time derivative inside the functional $\Delta L$ because of its linearity.
The positive parameter $|\dot{\Omega}_\infty|$ is settled by the intensity of the braking torque acting on the pulsar and represents the observed secular spin down of the pulsar.
Clearly, a completely analogous formula holds also in the Newtonian limit.


Considering that on the typical timescales involved in a glitch the right hand side in \eqref{eq:cons_ang_momentum} can be ignored, we define the \emph{maximal} glitch amplitude at a generic time $t$ as
\begin{equation}
\Delta \Omega_{\rm m}(t) \, =  \,  {\Delta L[\Omega_{np}(t)]} \, / \, I \, .
\label{eq:max_glitch}
\end{equation}
This quantity sets an upper limit to the amplitude of an hypothetical glitch that is triggered at time $t$, when the lag is $\Omega_{np}(t)$. It may be possible to produce an even larger glitch by allowing the lag $\Omega_{np}$ to become negative. This would produce a glitch overshoot, a fast transient phase that, according to current glitch simulations \citep{hasekll2012,graber+2018}, could occur within the first seconds after a large glitch is triggered. 
Recent analysis of a glitch in the Vela pulsar \citep{palfreyman2018Nat} points out that an overshoot is actually present in the timing data \citep{ashton2019NatAs,lettera_overshoot}. However, for most (if not all) of the data we use in the present analysis, this is not a problem, as the observed glitch size is likely to correspond to the jump in frequency at later times (see Fig 11 of \citealt{antonellipizzochero2017}). 

We now need a prescription to obtain $\Delta L[\Omega_{np}(t)]$. 
A way to proceed would be to employ a set of two-fluid hydrodynamic equations encoding macroscopic mutual friction \citep{AndSid06} and the effect of pinning \citep{seveso+2016}. Such equations would depend on the observed angular velocity $\Omega$ of the pulsar under study and on the inferred value of its secular spin down rate $\dot\Omega$. Moreover, the dynamical equations will also depend on some unknown structural properties of the star, like the EoS and the total mass, as well as on the parameters describing entrainment and pinning.

Once the theoretical curve $\Delta \Omega_{\rm m}(t)$ has been obtained, we still need to compare it with some information extracted from the observed timing behaviour of the particular pulsar under study.

\subsection{Contrasting the model with pulsar's timing data}
\label{susec:contrasting}

The quantity $\Delta \Omega_{\rm m}(t)$ sets a theoretical limit for the glitch amplitude at time $t$ in a pulsar that emptied its reservoir at $t=0$. 
However, we do not know when a pulsar actually empties its reservoir of angular momentum (maybe never). 

A sequence of maximal glitches, each emptying the reservoir, would result in a strong positive correlation between the glitch amplitudes and the waiting time between them, in contrast with the idea of glitches as random events that rarely empty the reservoir significantly \citep{melatos2008}. 
In such a system, the angular momentum released in each event is not expected to necessarily correlate with the angular momentum accumulated since the previous glitch: the effect of a finite-size reservoir, that can occasionally be emptied, is expected to generate only weak correlations between the glitch amplitude and the waiting time since the previous glitch \citep{MHF2018}. 
So far, these correlations induced by the finite size of the reservoir have not been observed in any pulsar, except only for the Vela at a low confidence level \citep{MHF2018}. 

Given the lack of evidence for backward waiting time-size correlation, the assumption that maximal glitches can occur in real pulsars may be satisfied only for very few events in some pulsars. Following Paper-I, we tentatively extend it to all pulsars showing large glitches, but only for their largest event in size. We denote by $\Delta\Omega_{\rm obs}$ the largest among the $\Delta\Omega_i$ observed glitches and assume that it corresponds to the total depletion of the available angular momentum reservoir.
We stress that there is no systematic argument for saying that the pulsar reaches corotation (i.e. it empties the reservoir) even during its largest observed glitch. On the contrary, it is expected that only a fraction of the accumulated angular momentum is released at each relaxation event in glitching puslars, but this is not a problem because this assumption is just used to put an upper bound to the mass. 

We now need to find a value for the typical timescale $t_{\rm act}$ between two events that may empty significantly the angular momentum reservoir. To do this we rely on an intrinsic property of the pulsar under study, the absolute activity $\mathcal{A}_a$.
Because of the random and impulsive nature of glitch sequences and of the slowness of the spin-down process (which implies low-number statistics), it can be difficult to extrapolate good estimates for $\mathcal{A}_a$ from glitch databases, except for a few pulsars (see Sec \ref{sec:sample}). 
For a pulsar which has undergone $N_{\rm gl}$ glitches of size $\Delta \Omega_i$ during an observational time interval $T$, the absolute activity could be estimated as
\begin{equation}
\mathcal{A}_a \approx \frac{1}{T} \sum_{i = 1}^{N_{\rm gl}} \Delta \Omega_i.
\label{eq:activity}
\end{equation}
We calculate this value by fitting the cumulative distribution of spin-up due to glitches \citep[see e.g. ][]{lyne2000}. In order not to overestimate the effect of the first and last glitch in the sequence, we perform a least-squares fit of the midpoints of the frequency jumps \citep{WongCrab}.
It is then possible to define the dimensionless activity $\mathcal{G}$ \citep{LE99} as 
\begin{equation}
\mathcal{G} \, = \,  \mathcal{A}_a\,  / \, |\dot{\Omega}| \, ,
\label{eq:G}
\end{equation}
that allows to compare pulsars of different spin-down rate. 

The activity has been employed several times to set an upper limit on the pulsar mass \citep{dattaalpar1993,LE99,andersson+2012, chamel2013}. 
The effect of different EoSs has also been studied, thus enabling to set observational upper limits on the mass \citep{delsate2016}. In this way it has been possible to test the average angular momentum reservoir associated to glitches. Nevertheless, there is no obvious dependence of $\mathcal{G}$ on the maximum glitch $\Delta\Omega_{\rm obs}$ observed for each object (see Table \ref{tab:pulsars}). 
The model proposed in Paper-I allows to account for both these parameters. 
This allows to partially solve the intrinsic degeneracy present in the definition of $\mathcal{A}_a$, namely the fact that we can obtain the same activity from several small glitches or from a few big ones.
From the activity and the largest observed glitch it is useful to define the characteristic time
\begin{equation}
 t_{\rm act} \, = \, \frac{\Delta\Omega_{\rm obs} }{ \mathcal{A}_a} \, = \, \frac{\Delta\Omega_{\rm obs}}{|\dot{\Omega}| \, \mathcal{G}} \, .
\label{eq:t_act}
\end{equation}
This represents the average inter-glitch time in an idealized object that has the same activity of the particular pulsar under study but follows a series of events of size $\Delta\Omega_{\rm obs}$.

To identity the \emph{single} glitchers (pulsars which in the observational time have displayed a single large glitch and several ones orders of magnitude smaller) we define the observational parameter
\begin{equation}
N_{\rm m} \, 
= \, \frac{\sum_{i = 1}^{N_{\rm gl}} \Delta \Omega_i}{\Delta\Omega_{\rm obs}} 
\, > \, 1 \, .
\label{eq:Nm}
\end{equation}
Single glitchers have $N_{\rm m} \approx 1$ and are not significative for the present analysis: at least two glitches of the same order of magnitude are necessary to give a rough estimate of $t_{\rm act}$. 
We interpret the smallness of $N_{\rm m}$ in single glitchers as an observational effect. As time goes by, these objects could eventually display another large glitch  and an activity estimate will then be more reliable. Clearly,
\begin{equation}
N_{\rm m} \, \approx \,  T \, / \, t_{\rm act}
\, ,
\label{eq:t_act_bis}
\end{equation}
meaning that $N_{\rm m}$ represents the number of events that the idealized pulsar would have displayed in the observational time. 
A large value for $N_{\rm m}$ indicates that $T$ has been long enough for the pulsar to potentially reach corotation several times: for this study, it is a better index of the statistical significance than $N_{\rm gl}$, the actual number of glitches detected during $T$ (see Sec \ref{sec:sample}).

We can now use the condition $\Delta\Omega_{\rm m}(t_{\rm act}) \geq \Delta\Omega_{\rm obs} $ to estimate $M_{\rm act}$. Since the largest observed glitch does not correspond to a complete exhaustion of the available reservoir, $M_{\rm act}$ is only an upper bound (a lighter star would still be compatible with the data).
The estimate $M_{\rm act}$ provides a refinement of the (less model dependent) absolute upper bound $M_{\rm max}$, given by emptying the fully-replenished reservoir compatible with pinning (as discussed in Paper-I and in the next section).


\section{Newtonian unified model}
\label{sec:model2}

Although the general form of the hydrodynamical equations is known, modelling mutual friction introduces some degree of arbitrariness, which is unavoidable due to the still poorly-understood vortex dynamics in neutron stars. 
The dynamical equations are therefore always phenomenological at some level, at least for what concerns aspects related to the unpinning and repinning of many vortices \citep{khomenko2018}.
For this reason, we now use the general concepts presented in the previous section by employing the particular toy-model presented in Paper-I. This model describes the rotational dynamics of a pulsar in a simplified way, but it captures the most important feature we are interested in: pulsars are slowly driven systems whose internal clock is set by the spin-down parameter $|\dot \Omega|$. 

For simplicity, we take the  Newtonian limit of Eqs \eqref{eq:ang_momentum} and \eqref{eq:cons_ang_momentum} and assume that the vortex lines are straight at the macroscopic scale\footnote{
	Cylindrical coordinates $(x, \varphi, z)$ are used, with $x$ representing the cylindrical radius, $\varphi$ the azimuthal angle and $z$ the coordinate along the rotation axis. The radius from the centre of the star is $r = \sqrt{x^2 + z^2}$.
	}. 
In this case, it is convenient to introduce the auxiliary variable 
\begin{equation}
\Omega_v \, = \, \Omega_p \, + \, (1 - \varepsilon_n) \Omega_{np} \, ,
\label{eq:omegav}
\end{equation}
where $\varepsilon_n(r)$ is the entrainment parameter \citep{prix2004}. 
In this way, the  rescaled lag 
\begin{equation}
\Omega_{vp} \, = \, \Omega_v - \Omega_p = (1 - \varepsilon_n) \Omega_{np}
\end{equation}
will depend on $x$ only, even if the entrainment parameter depends on $r$. 
The reservoir of angular momentum $\Delta L$ due to the presence of a lag $\Omega_{vp}$ turns out to be
\begin{equation}
\Delta L[\Omega_{vp}] = 2 \pi \int_0^R \mathrm{d}x\, x^3\, \Omega_{vp}(x) \int_{\gamma_x} \mathrm{d}l\, \frac{\rho_n(r)}{1 - \varepsilon_n(r)} \, ,
\label{eq:delta_l}
\end{equation}
where $R$ is the star radius, $\gamma_x$ is the curve that describes a straight vortex line placed at a distance $x$ from the rotation axis and $\rho_n(r)$ is the superfluid mass density.

Finally, the critical lag for the unpinning of vortices is obtained by equating the total Magnus and pinning forces along the line $\gamma_x$ \citep{antonellipizzochero2017}:
\begin{equation}
\Omega_{vp}^{\rm cr}(x) = \frac{\int_{\gamma_x} \mathrm{d}l\, f_P(r)}{\kappa\, x\, \int_{\gamma_x} \mathrm{d}l\, \frac{\rho_n(r)}{1 - \varepsilon_n(r)}},
\label{eq:crit_lag}
\end{equation}
where $\kappa = h / 2 m_n$ is the quantum of circulation and $f_P$ is the pinning force per unit length.

At this point, we could employ the set of two-fluid hydrodynamic equations described in \citet{antonellipizzochero2017}. As discussed in Sec. \ref{susec:following}, these equations should be solved for every pulsar with its distinctive values of $ \Omega$ and $\dot \Omega$.
We circumvent this complication by introducing a common unified timescale for pulsars with different spin down rates.
By taking as $t=0$ the moment in which $\Delta L=0$, we define the  nominal lag as $\omega^* \, = \, |\dot{\Omega}| \, t$. 
The increasing value of $\omega^*$ determines the actual rescaled lag built between the two components since corotation,
\begin{equation}
\Omega_{vp}(x,\, \omega^*) = \min{[\Omega_{vp}^{\rm cr}(x),\, \omega^*]}.
\label{eq:lag_t}
\end{equation} 
%
According to the general approach outlined in Sec. \ref{susec:contrasting}, the typical nominal lag elapsed between two large glitches is 
\begin{equation}
\omega^*_{\rm act} = t_{\rm act} |\dot{\Omega}| 
\, = \,  \Delta\Omega_{\rm obs} \, / \,  \mathcal{G} \, .
\label{eq:omega_act}
\end{equation}
If we measure a maximum glitch amplitude $\Delta \Omega_{\rm obs}$ for a particular pulsar of known activity, we can invert the relation $\Delta \Omega_{\rm m} (\omega^*_{\rm act}, M_{\rm act}) = \Delta \Omega_{\rm obs}$ to obtain $M_{\rm act}$.

Up to this point, we have not assumed anything about the location and extension of the region in which the neutron superfluid resides, i.e. the region in which $\rho_n >0$. In the case of the maximum glitch amplitude, corresponding to the critical lag in Eq.~\eqref{eq:crit_lag}, we have
\begin{equation}
\Delta \Omega_{\rm max} \, = \, 
\Delta \Omega_{\rm m}(t\rightarrow \infty) \, = \,
\Delta L[ \Omega_{vp}^{\rm cr}(x) ]\, / \, I \, . 
\label{eq:maximum_glitch0}
\end{equation}
It can be shown that 
\begin{equation}
\Delta \Omega_{\rm max} = \frac{\pi^2}{I \kappa} \int_0^{R_d} \mathrm{d} r\, r^3\, f_P(r)\, ,
\label{eq:maximum_glitch}
\end{equation}
where $R_d$ is the neutron-drip radius (the outer edge of the inner crust, at baryon density $n_d=2.6\times 10^{-4}$ fm$^{-3}$). 
Note that $\Delta \Omega_{\rm max}$ depends only on the extension of the pinning region. 
If $f_P$ is non-zero only in the inner crust (i.e. for $R_c < r < R_d$, where $R_c$ is the crust-core boundary), the integral in Eq. \eqref{eq:maximum_glitch} receives no contribution from the core. 
Therefore, $\Delta \Omega_{\rm max}$ does not depend on the vortex extension, provided that they extend at least in the pinning region of the crust of the star and that pinning in the core is negligible. 
On the contrary, the maximal glitch amplitude $\Delta \Omega_{\rm m}$ is different according to the region where we assume the presence of the superfluid, due to the explicit dependence on $\rho_n(r)$ in Eqs.~\eqref{eq:delta_l} and \eqref{eq:crit_lag}: 
considering the superfluid limited to spherical shells ending at different depths in the core changes the value of $\Delta \Omega_{\rm m} (\omega^*, M)$, and therefore the estimate of $M_{\rm act}$.


\section{Pulsar sample}
\label{sec:sample}

\begin{figure}
	\centering
	\includegraphics[width = 0.47 \textwidth]{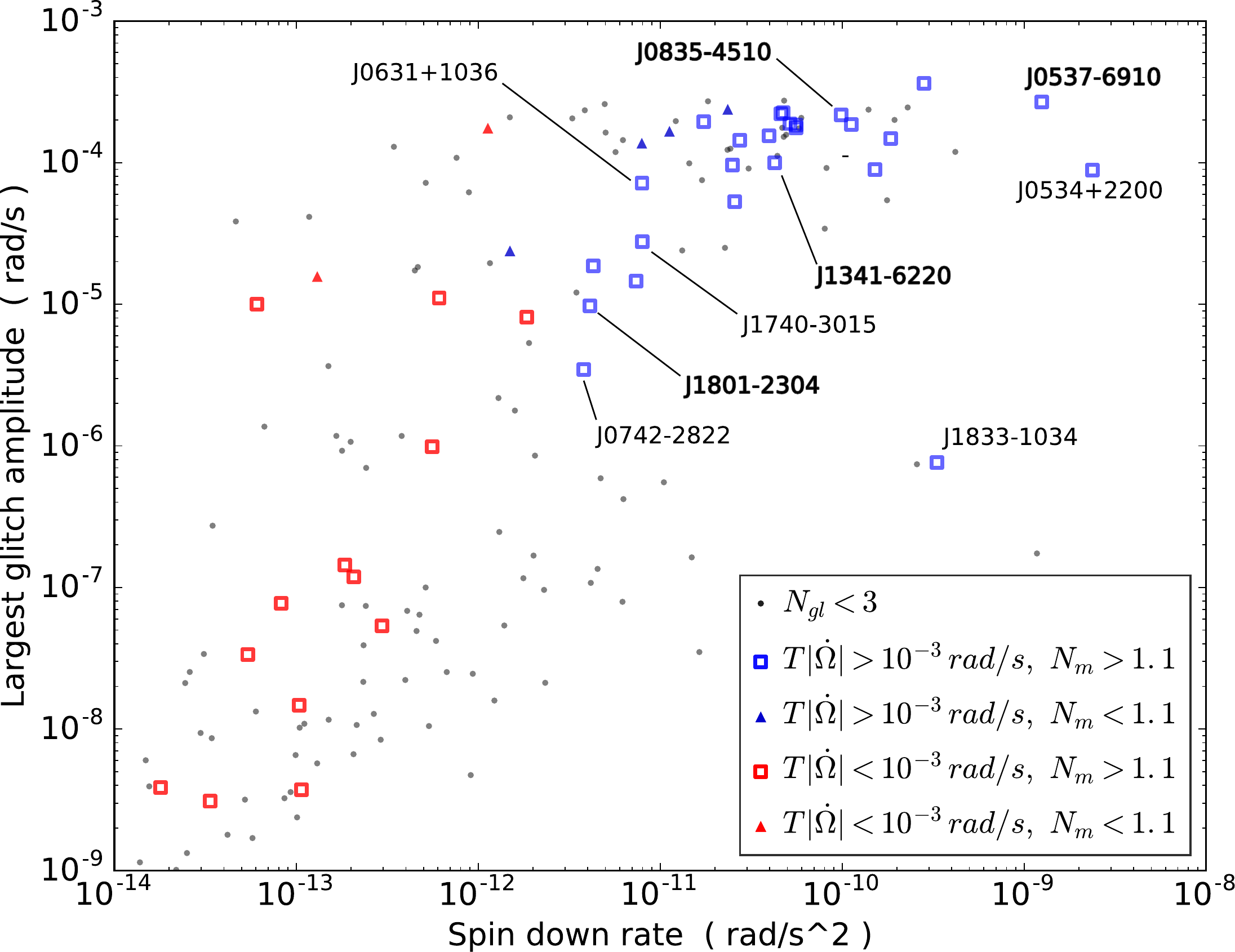}
	\caption{The largest glitch amplitude $\Delta\Omega_{\rm obs}$ observed in 166 glitching pulsars as a function of their spin-down rate, $ |\dot{\Omega}|$. We indicate by grey points the pulsars with $N_{\rm gl} < 3 $, by triangles the single glitchers with $N_{\rm m} \leq 1.1 $, by squares the remaining objects with $N_{\rm gl} \geq 3 $ and $N_{\rm m} > 1.1 $. The four pulsars indicated with a bold name have $N_{\rm m} > 4 $. The squares and triangles are displayed in red if $ T |\dot{\Omega}| < 10^{-3}$ rad/s and in blue if $ T |\dot{\Omega}| > 10^{-3}$ rad/s.
	}
	\label{fig:sample}
\end{figure}
\begin{figure}
	\centering
	\includegraphics[width = 0.47 \textwidth]{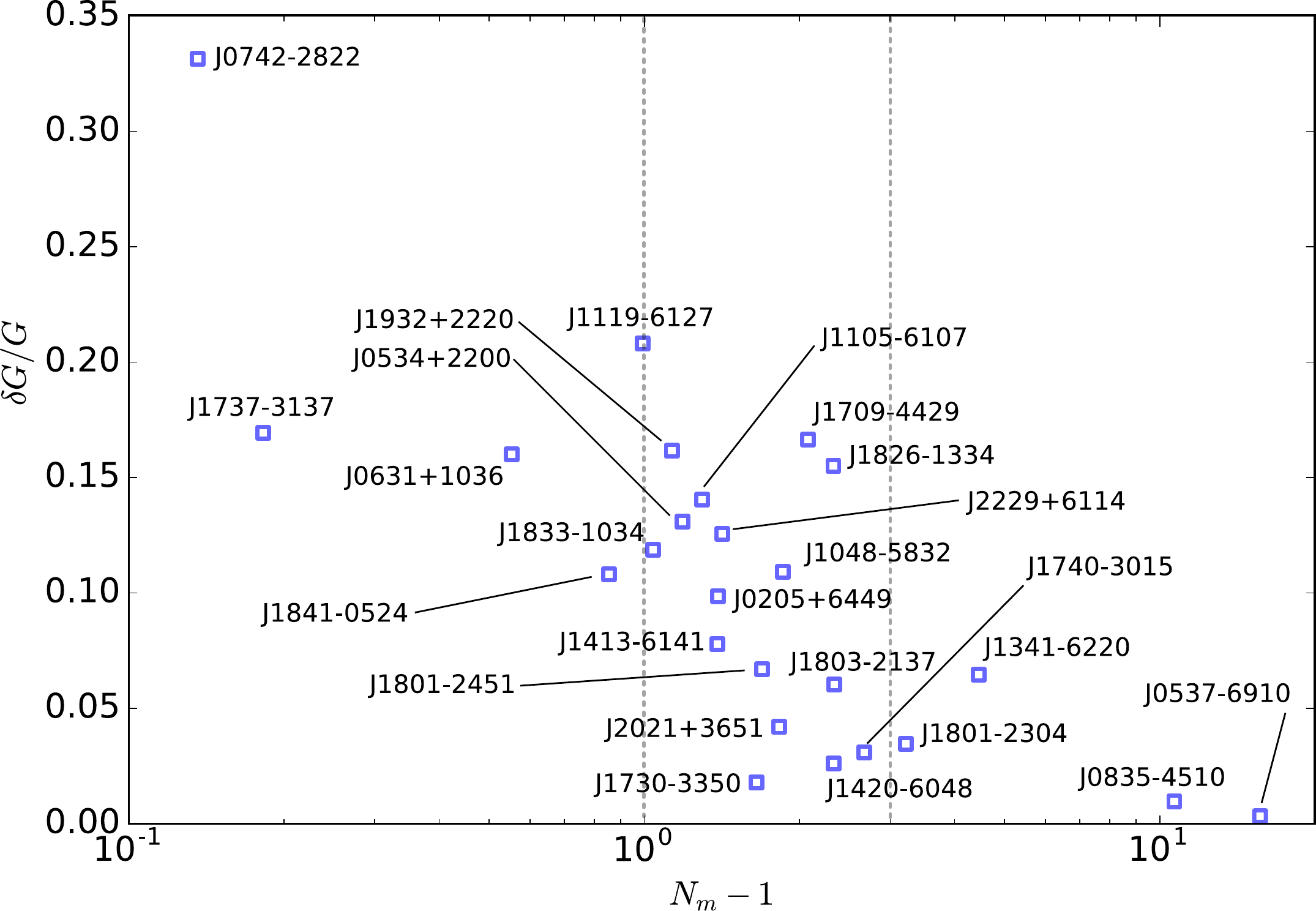}
	\caption{The relative error on the activity parameter, $\delta \mathcal{G} / \mathcal{G}$, as a function of $(N_{\rm m} -1)$ for the 25 pulsars in our sample (see Table~\ref{tab:pulsars}). The vertical lines mark the $N_{\rm m}=2$ and $N_{\rm m}=4$ boundaries.
	}
	\label{fig:Nmax}
\end{figure}

In this work we select a sample from 166 known glitching pulsars, by crossing the information obtained from the Jodrell Bank Glitch Catalogue \citep[\url{www.jb.man.ac.uk/pulsar/glitches.html}, ][]{espinoza+2011} and from the ATNF Pulsar Catalogue \citep[\url{www.atnf.csiro.au/research/pulsar/psrcat}, ][]{manchester+2005}. 
The sample should contain glitchers statistically relevant for our approach, namely pulsars whose activity can be determined and which are not too affected by observational biases, such as a short observational time $T$.

In Fig~\ref{fig:sample} we display the largest observed glitch, $\Delta\Omega_{\rm obs}$, for the 166 glitchers as a function of the pulsar's spin-down rate, $|\dot \Omega|$. 
In the figure, we indicate by points the objects with $N_{\rm gl}<3 $: we eliminate these pulsars from the sample, since at least 3 glitches are needed (but not sufficient) to try to give a rough fit of the activity of the star. Then, we indicate by triangles the single glitchers (defined by $N_{\rm m} \leq 1.1$): as discussed in Sec. 2.2, a reasonable activity cannot be determined with a single large event and thus these objects are also excluded from the sample. Finally, the remaining glitchers are indicated by squares: for these objects, an estimate of their activity can be obtained.

Looking at the square symbols in Fig~\ref{fig:sample}, it can be seen that fast evolving pulsars (of large $|\dot{\Omega}|$) exhibit large maximum events, while slowly evolving ones display only small values of $\Delta\Omega_{\rm obs}$, as already reported in previous works \citep{McKenna1990Nat,lyne2000,espinoza+2011,yu2013}. 
Note that these tiny events, with $\Delta\Omega_{\rm obs} < 10^{-6}$ rad/s, may be due to a different glitch mechanism, not hidden by the more unfrequent larger glitches  \citep{fuentes17,ashton2017}. 
This is probably an observational effect, as slowly evolving pulsars need more time to develop a lag sufficient for a large glitch, thus they need to be monitored for a longer interval $T$.
To quantify the effect of the observational time $T$, we introduce the observational nominal lag $T\, |\dot{\Omega}| $. This quantity represents the maximum lag that could have been developed since the pulsar has been observed, and allows for comparison of different objects. 
Since the typical $ \omega^*_{\rm act} $ is always larger than $10^{-3}$ rad/s (see Fig \ref{fig:sample} and Tab \ref{tab:pulsars}), it is reasonable to require that pulsars in the sample must have been observed long enough to develop at least such a lag. In the figure, we denote by red symbols the glitchers with $T\, |\dot{\Omega}| < 10^{-3}$ rad/s, and by blue symbols those with $ T\, |\dot{\Omega}| > 10^{-3}$ rad/s and use this criterion to distinguish the fast evolving pulsars from the slowly evolving ones.

Summarising, we select our sample by requiring three specific conditions:
\begin{itemize}
	\item $N_{\rm gl} \geq 3 $ - This is required to to fit the activity.
	\item $N_{\rm m} >1.1 $ - To eliminate the single glitchers from the sample. The low threshold 1.1 has been chosen to select, as a first tentative step, a large number of potentially interesting objects with diverse rotational parameters and glitch amplitudes. Changing the threshold to $N_{\rm m} > 1.5$, however, would only remove two objects from the sample. 
	\item $ T\, |\dot{\Omega}| > 10^{-3}$ rad/s - To eliminate the pulsars that evolve slowly (and so require a lot of time to replenish the reservoir) or that have not been observed for a sufficiently long period $T$.
\end{itemize} 
In this way, we obtain the sample of 25 stars in Table~\ref{tab:pulsars}. The pulsars studied in Paper-I are all present in the sample, plus eight additional objects. Since publication of Paper-I, the glitch sequence of PSR J0537-6910 \citep{antonJ0537} was re-analysed, yielding smaller uncertainties on the observed glitch amplitudes: 
for this reason, the mass estimate of this pulsar has very small errors in the present work.
\begin{table*}
	\centering
	\begin{tabular}{ccccccccc@{}}
		\hline
		  J-name   & $\Omega$ & $|\dot{\Omega}|$ &   $\mathcal{G}$   & $N_{\rm gl}$ & $N_{\rm m}$ & $\Delta \Omega_{\rm obs}$ & $T\, |\dot{\Omega}|$ & $\omega^*_{\rm act}$ \\
		           & [rad/s]  & [$10^{-11}$ rad/s$^2]$  &      $[\%]$       &              &               &     $[10^{-4}$ rad/s]     &           [$10^{-3}$ rad/s]           &  [$10^{-3}$ rad/s]   \\ \hline\hline
		J0205+6449 &  95.61   &          28.19          & 0.743 $\pm$ 0.073 &      6       &     2.39      &     3.633 $\pm$ 0.382     &                104.2                 &   48.90 $\pm$ 7.05   \\
		J0534+2200 &  188.2  &         237.2          &$(4.0\pm0.5)\times 10^{-3}$&      27      &     2.19      &     0.886 $\pm$ 0.006     &         3636                & 2232 $\pm$ 292 \\
		J0537-6910 &  389.7  &         125.2          & 0.874 $\pm$ 0.003 &      45      &     16.66     &     2.677 $\pm$ 0.012     &                491.1                 &   30.62 $\pm$ 0.17   \\
		J0631+1036 &  21.83   &          0.79           & 1.333 $\pm$ 0.213 &      15      &     1.55      &           0.716           &                 3.79                  &   5.37 $\pm$ 0.86    \\
		J0742-2822 &  37.68   &          0.38           & 0.107 $\pm$ 0.036 &      8       &     1.14      &     0.035 $\pm$ 0.001     &                 2.99                  &   3.23 $\pm$ 1.07    \\ \hline
		J0835-4510 &  70.34   &          9.84           & 1.616 $\pm$ 0.016 &      20      &     11.67     &     2.180 $\pm$ 0.008     &                148.4                 &   13.49 $\pm$ 0.14   \\
		J1048-5832 &  50.81   &          3.96           & 1.623 $\pm$ 0.177 &      6       &     2.86      &           1.546           &                 18.98                 &   9.53 $\pm$ 1.04    \\
		J1105-6107 &  99.43   &          2.49           & 1.311 $\pm$ 0.184 &      5       &     2.30      &           0.966           &                 10.48                 &   7.37 $\pm$ 1.04    \\
		J1119-6127 &  15.40   &          15.18          & 0.175 $\pm$ 0.036 &      4       &     1.99      &     0.892 $\pm$ 0.031     &                 81.29                 &  50.91 $\pm$ 10.74   \\
		J1341-6220 &  32.50   &          4.25           & 1.524 $\pm$ 0.098 &      23      &     5.46      &           1.000           &                 26.09                 &   6.56 $\pm$ 0.42    \\ \hline
		J1413-6141 &  22.00   &          2.57           & 1.375 $\pm$ 0.107 &      7       &     2.38      &           0.530           &                 6.65                  &   3.86 $\pm$ 0.30    \\
		J1420-6048 &  92.16   &          11.24          & 1.366 $\pm$ 0.036 &      5       &     3.33      &     1.861 $\pm$ 0.012     &                 37.00                 &   13.62 $\pm$ 0.37   \\
		J1709-4429 &  61.32   &          5.57           & 1.389 $\pm$ 0.231 &      4       &     3.08      &     1.761 $\pm$ 0.016     &                 28.54                 &   12.68 $\pm$ 2.11   \\
		J1730-3350 &  45.05   &          2.74           & 1.403 $\pm$ 0.025 &      3       &     2.65      &           1.443           &                 18.78                 &   10.28 $\pm$ 0.19   \\
		J1737-3137 &  13.95   &          0.43           & 1.144 $\pm$ 0.194 &      4       &     1.18      &           0.187           &                 2.06                  &   1.64 $\pm$ 0.28    \\ \hline
		J1740-3015 &  10.35   &          0.80           & 1.216 $\pm$ 0.038 &      36      &     3.67      &           0.276           &                 7.73                  &   2.27 $\pm$ 0.07    \\
		J1801-2304 &  15.11   &          0.41           & 1.009 $\pm$ 0.035 &      13      &     4.22      &           0.098           &                 3.77                  &   0.97 $\pm$ 0.03    \\
		J1801-2451 &  50.30   &          5.15           & 1.720 $\pm$ 0.115 &      5       &     2.69      &           1.889           &                 23.07                 &   10.98 $\pm$ 0.73   \\
		J1803-2137 &  47.01   &          4.72           & 1.781 $\pm$ 0.107 &      5       &     3.34      &     2.253 $\pm$ 0.001     &                 30.74                 &   12.65 $\pm$ 0.76   \\
		J1826-1334 &  61.91   &          4.59           & 1.281 $\pm$ 0.199 &      6       &     3.33      &     2.217 $\pm$ 0.001     &                 39.86                 &   17.31 $\pm$ 2.68   \\ \hline
		J1833-1034 &  101.5  &          33.14    &$(3.6\pm0.4)\times 10^{-3}$   &      4       &     2.04      &           0.008           &            27.86                 &   21.27 $\pm$ 2.54   \\
		J1841-0524 &  14.10   &          0.74           & 1.532 $\pm$ 0.166 &      5       &     1.85      &           0.145           &                 1.92                  &   0.95 $\pm$ 0.10    \\
		J1932+2220 &  43.49   &          1.73           & 4.513 $\pm$ 0.729 &      3       &     2.13      &           1.945           &                 6.39                  &   4.31 $\pm$ 0.70    \\
		J2021+3651 &  60.57   &          5.59           & 1.609 $\pm$ 0.067 &      4       &     2.82      &     1.846 $\pm$ 0.001     &                 22.09                 &   11.47 $\pm$ 0.48   \\
		J2229+6114 &  121.7  &          18.45          & 0.522 $\pm$ 0.066 &      6       &     2.42      &     1.487 $\pm$ 0.005     &                 52.66                 &   28.48 $\pm$ 3.58   \\ \hline
	\end{tabular}
	\caption{The sample of pulsars used in this work. No errors are reported when they are smaller than the symbols used in figures. The timing data and their observational uncertainties have been obtained by crossing the information from the ATNF Pulsar Catalogue (\url{www.atnf.csiro.au/research/pulsar/psrcat}, see also \citealt{manchester+2005}) and the Jodrell Bank Glitch Catalogue (\url{www.jb.man.ac.uk/pulsar/glitches.html}, see also \citealt{espinoza+2011}).}
	\label{tab:pulsars}
\end{table*}
\\
In Fig \ref{fig:Nmax} we plot the relative error on the activity parameter, $\delta \mathcal{G} / \mathcal{G}$, as a function of $(N_{\rm m}-1)$. 
We see that the error is larger than the $10\%$ for $N_{\rm m}<2$ and smaller than $7\%$ for $N_{\rm m}>4$ (the vertical lines at 2 and 4 are drawn for visual clarity). 
This shows how larger values of the parameter $N_{\rm m}$ are associated to more precise estimates of the activity and hence indicate the statistical significance of the glitch sequence, as it has been anticipated in Sec \ref{susec:contrasting}. 
The same is not true for the total number of glitches. For example, the Crab and PSR J0631+1036 have displayed a large number of glitches ($N_{\rm gl}=27$ and $15$ respectively), but they show low values of $N_{\rm m}$ ($=2.19$ and $1.55$ respectively) and their activities have errors larger than $10\%$.
Hence, in the following we take the four pulsars with $N_{\rm m}>4$ as benchmarks for the study of the mass of pulsars with very different glitch size.


\section{Results}
\label{sec:results}

We now study the dependence of the mass estimate $M_{\rm act}$ on the extension of the superfluid reservoir. 
To do so, we perform different spherical cutoffs in the extension of the superfluid region involved in the glitch, by imposing that the reservoir extends from neutron drip density to $1\,n_0$, $0.75\,n_0$, $0.68\,n_0$ and $0.6\,n_0$, where $n_0 = 0.168\, \mbox{fm}^{-3}$ is the nuclear saturation density \citep{chamelReview}. 
Finally, we consider a superfluid reservoir limited to the crust (where the crust-core boundary, $n_c$, is given by the specific EoS implemented).
The choice of these cutoffs is justified by physical motivations: the region between the crust-core interface and $1\,n_0$ is the region where most of the theoretical superfluid gaps of singlet state ${}^1S_0$ go to zero. In particular, $0.68\, n_0$ corresponds to the value where the superfluid region ends in a neutron star with temperature $T \approx 10^8 K$, considering a SFB superfluid gap \citep{schwenk+2003,ho+2015}.

We consider two unified EoSs, SLy4 \citep{douchinhaensel2001}, BSk20 \citep{goriely+2010}, and a stiffer relativistic mean field model, DDME2 \citep{LNV2005}, see  Tab~\ref{tab:eos}.  
The DDME2 EoS does not have any consistently calculated superfluid neutron fraction $x_n$ in the crust, so that we glued it with the crust from the SLy4 EoS, keeping the crust-core transition density to be the one of SLy4. 
This operation has been carried out by ensuring the continuity of the chemical potential, as discussed by \cite{fortinC16}. This, while ensuring thermodynamic consistency, also produces a strong first-order phase transition at the crust-core interface: the $P(n_b)$ profile of the DDME2+SLy4 EoS turns out to be flat for $n_b$ between $ 0.076\,$fm$^{-3} $ and $ 0.084\,$fm$^{-3}$ (namely, $0.45 \, n_0$ and $ 0.5 \,n_0$), and a corresponding density jump appears at the crust-core interface. 

\begin{table}
	\centering
\setlength{\tabcolsep}{15pt}
	\begin{tabular}{@{}cccccccccc@{}}
		\hline
		EoS   & $M_{\rm eos}$  $[M_{\odot}]$& $n_c$ [$n_0]$\\ \hline \hline
        SLy4  & 2.05          & 0.452\\
        BSk20 & 2.16          & 0.508\\
        DDME2 + SLy4 & 2.48   & 0.452\\ \hline
	\end{tabular}
	\caption{The maximum neutron star mass and the baryon density corresponding to the crust-core transition
	for the EoSs used in this work.}
	\label{tab:eos}
\end{table}

\begin{figure}
\centering
\includegraphics[width = \columnwidth]{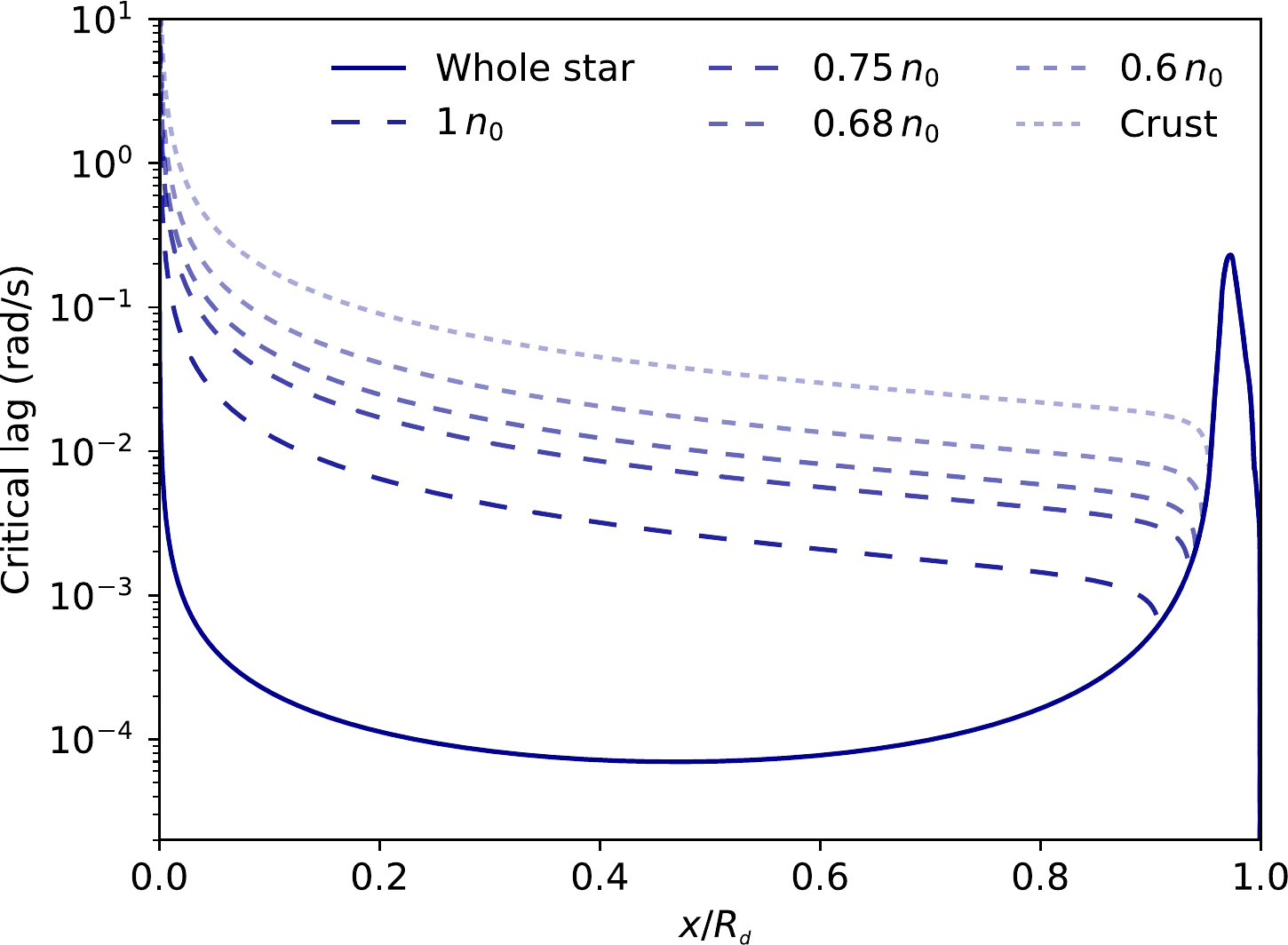}
\caption{ Example of critical lag $\Omega_{vp}^{\rm cr}$, as a function of the radius $x/R_d$ and for the different cutoff densities $n_{\rm cut}$. The calculation refers to a $1.4\, M_\odot$ star, described by the BSk20 EoS. 
	}
\label{fig:crit_lag}
\end{figure}

In Fig~\ref{fig:crit_lag} we show the critical lag for straight vortex lines, given by Eq.~\eqref{eq:crit_lag}, for the different cutoffs considered here. 
The calculation was done with the BSk20 EoS, by employing the pinning force of \citet{seveso+2016} and the entrainment parameters obtained in \citet{chamelhaensel2006} for the core and \citet{chamel2012} for the crust. 
As expected, $\Omega_{vp}^{\rm cr}$ has higher values in the central region of the star for smaller superfluid reservoirs: since the superfluid extends in a smaller spherical layer, the superfluid vortices are less subject to the Magnus force.
Since the critical lag is   cutoff dependent, the lag $\Omega_{vp}$ (and hence $\Delta \Omega_{\rm m}$, via Eq.~\eqref{eq:max_glitch}) evolves differently. However, $\Delta \Omega_{\rm max}$ does not depend on the cutoff we are considering, as the different form of the critical lag is compensated by the second integral over $\gamma_x$ in Eq.~\eqref{eq:delta_l}.

\begin{figure*}
\centering
\includegraphics[width = \textwidth]{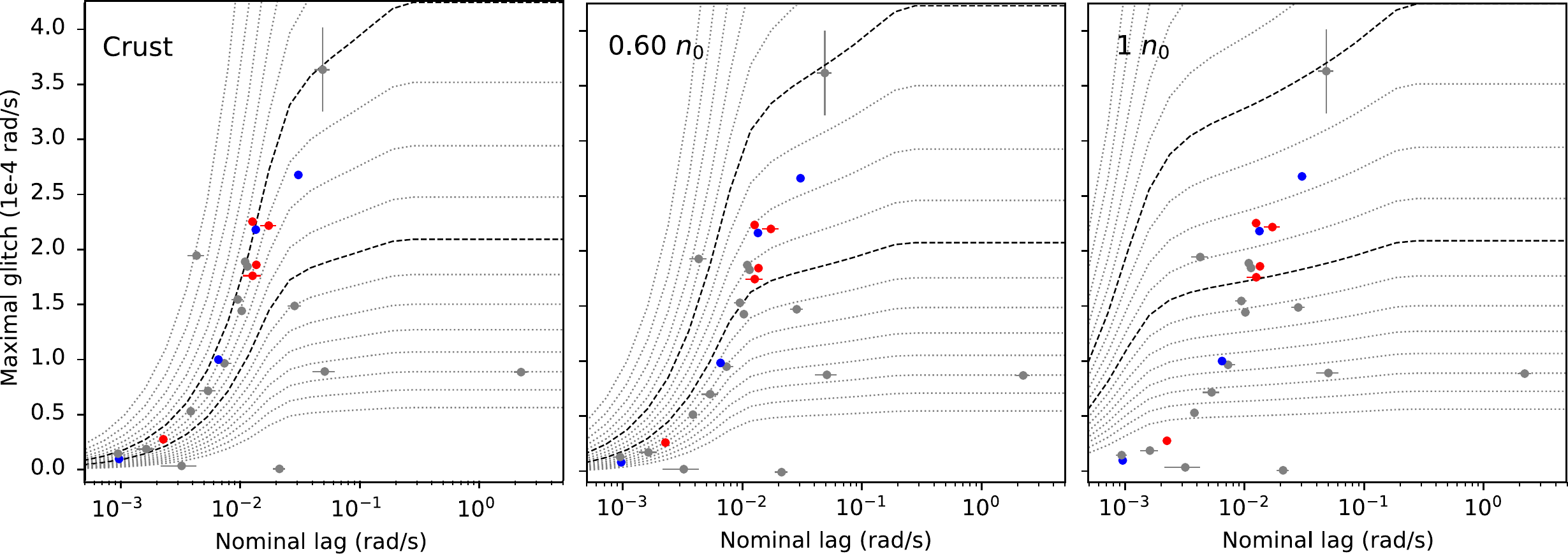}
\caption{
The maximal glitch $\Delta \Omega_{\rm m}(\omega^*, M, n_{\rm cut})$, as a function of the nominal lag $\omega^*$, for different pulsar masses $M$ and reservoir cutoffs $n_{\rm cut}$, in the case of the BSk20 EoS. 
The curves in each panel correspond to different masses, starting from $0.5 M_\odot$ (the highest curve) up to $2.1 M_\odot$ (the lowest one); the $1.0 M_\odot$ and the $1.4 M_\odot$ curves are highlighted (black dashed lines).
We also display the values of the largest observed glitch $\Delta \Omega_{\rm obs}$ and the nominal lag $\omega_{\rm act}^*$ for the 25 pulsars in Tab~\ref{tab:pulsars}.  
Pulsars with $N_{\rm m} < 3$ are shown in grey, the ones with $3 < N_{\rm m} < 4$ in red, and the remaining ones with $N_{\rm m} > 4$ in blue. 
}
\label{fig:omega_act}
\end{figure*}

We now study the evolution of the maximal glitch, $\Delta \Omega_{\rm m}(\omega^*)$, as a function of the nominal lag. The results are shown in Fig~\ref{fig:omega_act}, where it can be seen that the maximal glitch raises faster as a function of $\omega^*$ for more extended reservoirs, in particular for lower masses. 
On the other hand, for large values of $\omega^*$ the maximal glitch tends to $\Delta \Omega_{\rm max}$, which in the present scenario of crustal pinning does not depend on the superfluid cut.
The stars of the sample seem to follow the form of the curves for the masses, especially in the case of a reservoir limited to the crust; this may be just a coincidence related to the fact that most pulsars of our sample are aligned along $\mathcal{G}\sim 1 \% $, as also pointed out by \citealt{fuentes17}.
As a consequence of this fact, the mass estimates for the crust-limited reservoir will fall in a narrow range of values.

We notice that two objects (J0742-2822 and J1833-1034) are below the lowest curve, corresponding to the highest mass achievable from BSk20: they are not constrained by the reservoir, in the sense that any mass compatible with the EoS could yield such small glitches (with $\Delta \Omega_{\rm obs} < 5\times 10^{-6}$ rad/s). Interestingly, this value is just below the dividing line $\Delta\Omega/2 \pi \approx 10\, \mu$Hz found by \citet{espinoza+2011} by analysing the bimodal distribution of all measured glitch sizes. Hence, these unconstrained objects may belong to a subpopulation which is unable to release a sufficient amount of angular momentum to produce large glitches (cf. also the more recent analysis of \cite{ashton2017} and \cite{fuentes17}). 

Another viable hypothesis is that J0742-2822 and J1833-1034 have not displayed yet a glitch large enough.
In fact, these pulsars still have small values of $N_{\rm m} \lesssim 2$, a case that resembles the Crab (J0534+2200, $N_{\rm m}=2.2$), which only after 50 years of observations has displayed a glitch big enough to be relevant for the present analysis \citep{shaw+2018}. 
The very large value of $\omega_{\rm act}^*$ associated to the Crab may be due to its young age and possible thermal effects favoured by high temperatures (e.g. enhanced vortex creep, implying longer times to build up the excess of angular momentum). 


Another peculiar object is PSR J1932+2220, with its low value of $\omega_{\rm act}^*$. In the crust-limited reservoir case, it is marginally fitted by the low $ 0.5\, M_\odot$ curve. However, we also notice that in the $0.68\, n_0$ case the star is well within the 1-1.4 $M_\odot$ region. Thus, for this star the superfluid reservoir should be extended to a small region in the outer core to obtain reasonable masses; future observations and improved statistics may change the situation (this pulsar has low values of both $N_{\rm m}=2.1$ and $N_{\rm gl}=3$). 

\begin{figure*}
\centering
\includegraphics[width = \textwidth]{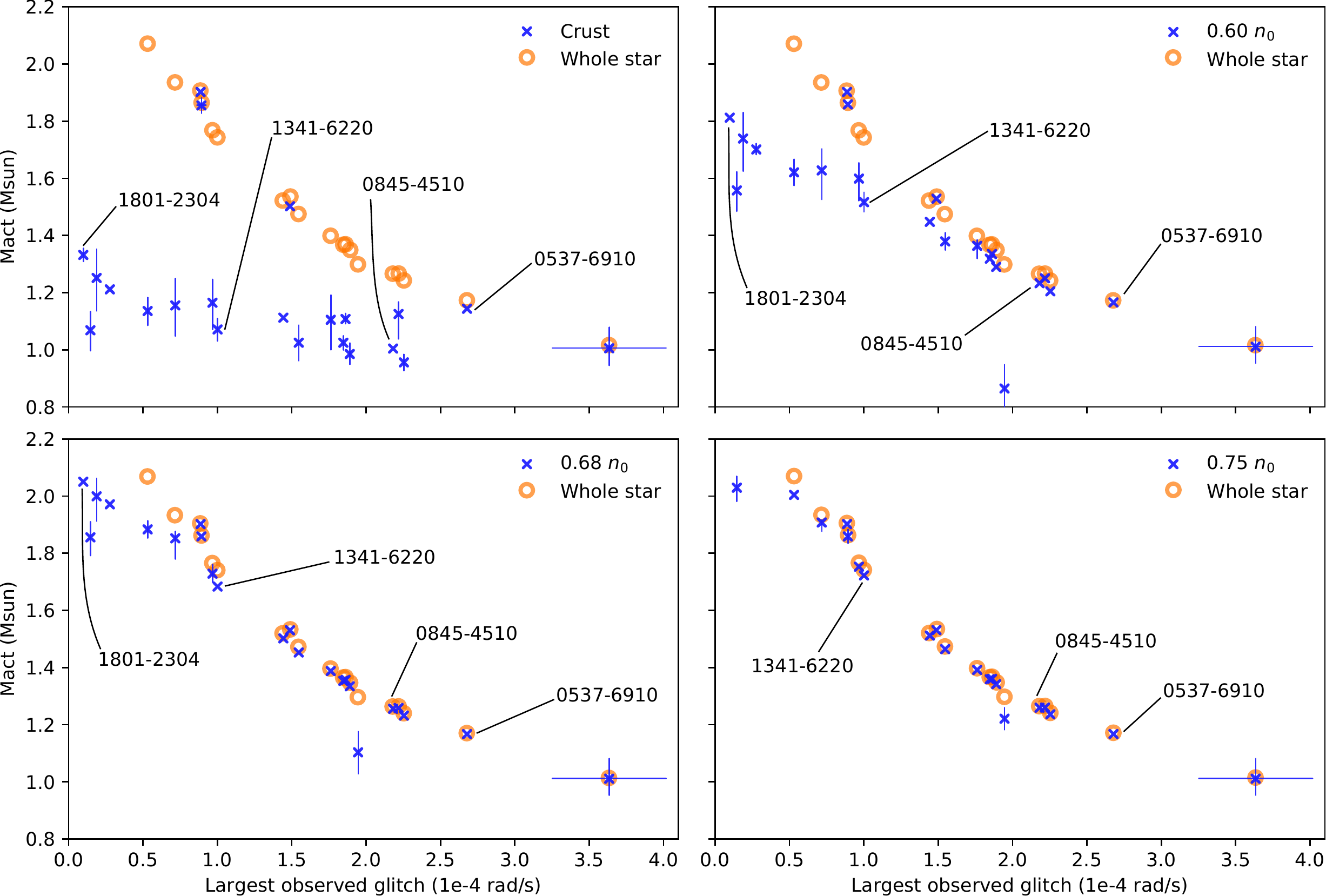}
\caption{
The 25 fitted values $M_{\rm act}$ as a function of $\Delta \Omega_{\rm obs}$ in the case of the BSk20 equation of state. Each panel refers to a different cutoff $n_{\rm cut}$ (blue crosses). Error bars are absent when smaller than the symbols used. For comparison, in every panel we also show the case of the reservoir extending to the whole star (orange circles, error bars not displayed). The $n_{\rm cut} =1\, n_0$ cutoff is omitted, as it is almost identical to the whole star case. The four pulsars with $N_{\rm m} > 4$ are also indicated.	}
\label{fig:masses}
\end{figure*}

For a given cutoff $n_{\rm cut}$, we invert the equation $\Delta \Omega_{\rm m} (\omega^*_{\rm act}, M_{\rm act}) = \Delta \Omega_{\rm obs}$ for each pulsar and find the value of $M_{\rm act}$. Fig \ref{fig:omega_act} provides a graphical representation of this procedure. 
The results for the BSk20 EoS are shown in Fig~\ref{fig:masses}, where we plot the mass estimate $M_{\rm act}$ as a function of the largest observed glitch $\Delta \Omega_{\rm obs}$. 
In each panel, we show the masses corresponding to a particular cutoff, and give as a reference the case of no-cutoff (reservoir extended to the entire star, the case previously considered in Paper-I). 
The cutoff at $1\,n_0$ has been omitted, since the corresponding mass estimates are identical to the case of the whole star. Thus, there is no particular need to invoke inner parts of  the core (where $n_b>n_0$) to explain current glitch data. 
This is good news, considering the present uncertainty of theoretical calculations on the properties P-wave pairing gap in the core and the presence of a layer of normal matter between the triplet and singlet neutron superfluids. 

It is possible to notice some general trends in our results. 
First, in Fig~\ref{fig:masses} we see an inverse correlation between amplitude of the largest glitch and estimated mass. Clearly, the inverse relation between the maximum theoretical glitch amplitude $\Delta\Omega_{\rm max}$ in \eqref{eq:maximum_glitch0} and the mass is obvious and expected, as lighter stars have thicker crusts.
However, this inverse relation persists also when we use the maximal amplitude $\Delta\Omega_{\rm m}$ defined in \eqref{eq:max_glitch}: in this case the anti-correlation between the estimated mass and the largest glitch amplitude observed is less obvious because the parameter $t$ in \eqref{eq:max_glitch}, which we estimated from the observed glitch activity via $t \approx \omega^*_{\rm act} / |\dot{\Omega}|$, is different for every pulsar.

The slope of the curves in Fig~\ref{fig:masses} increases with increasing extension of the reservoir, being almost flat for the crust-only case and tending to the whole-star case already for $n_{\rm cut} \gtrsim 1\, n_0$. Also, if we extend the superfluid reservoir to deeper regions of the star we can fit less masses than in the case of a smaller reservoir: in Fig~\ref{fig:omega_act} some pulsars with small largest glitch and small nominal lag can only be constrained in the cases of more external cutoffs. 

Secondly, objects with a small nominal lag ($\omega_{\rm act}^* \lesssim 2 \times 10^{-2}$ rad/s) are more sensitive to changes of the cutoff than those with a large one. In fact, pulsars with small nominal lag show masses around 1.0-$1.4\, M_\odot$ in the case of reservoir limited to the crust, while they show much larger masses (or they do not even get constrained) for more extended cutoffs. 
On the other hand, the five pulsars with the largest nominal lag ($\omega_{\rm act}^* \gtrsim 2.5 \times 10^{-2}$ rad/s) have their masses almost unaltered between the different cutoffs, as can be noticed in Fig~\ref{fig:masses}. 
The reason for this is easy to understand in the case of the Crab, with its extreme value of the nominal lag: when $\omega^*$ is large enough, the lag as a function of time \eqref{eq:lag_t} has reached the critical value \eqref{eq:crit_lag}. 
As a consequence, the maximal glitch reaches a plateau, given by the maximum glitch amplitude \eqref{eq:maximum_glitch}. In fact, for pulsars with large $\omega_{\rm act}^*$, the maximal glitch corresponds to the maximum glitch: their mass estimates are independent on the superfluid reservoir extension or entrainment parameters, but strongly dependent on the pinning force considered. 
Although the four remaining pulsars with large nominal lag (among which two other peculiar objects, J1119-6127 and J2229+6114) have not yet reached the plateau, they still lie in a region of the $(\omega^*, \, \Delta\Omega_{\rm obs})$ plane where the curves $\Delta \Omega_{\rm m}(\omega^*)$ are almost insensitive to the choice of the cutoff.

It is interesting to notice how, for the crust-limited reservoir, the masses of the pulsar are - except for the three pulsars with the largest $\omega_{\rm act}^*$ (Crab, J1119-6127 and J2229+6114) - all quite low, peaked around $\approx 1.1\, M_\odot$ and even less than $\approx 1\, M_\odot$ in some cases. This fact indicates that the crustal reservoir alone is not enough to describe pulsar glitches, as already noticed by \citet{andersson+2012} and \citet{chamel2013}. 


\begin{figure*}
\centering
\includegraphics[width = 0.9 \textwidth]{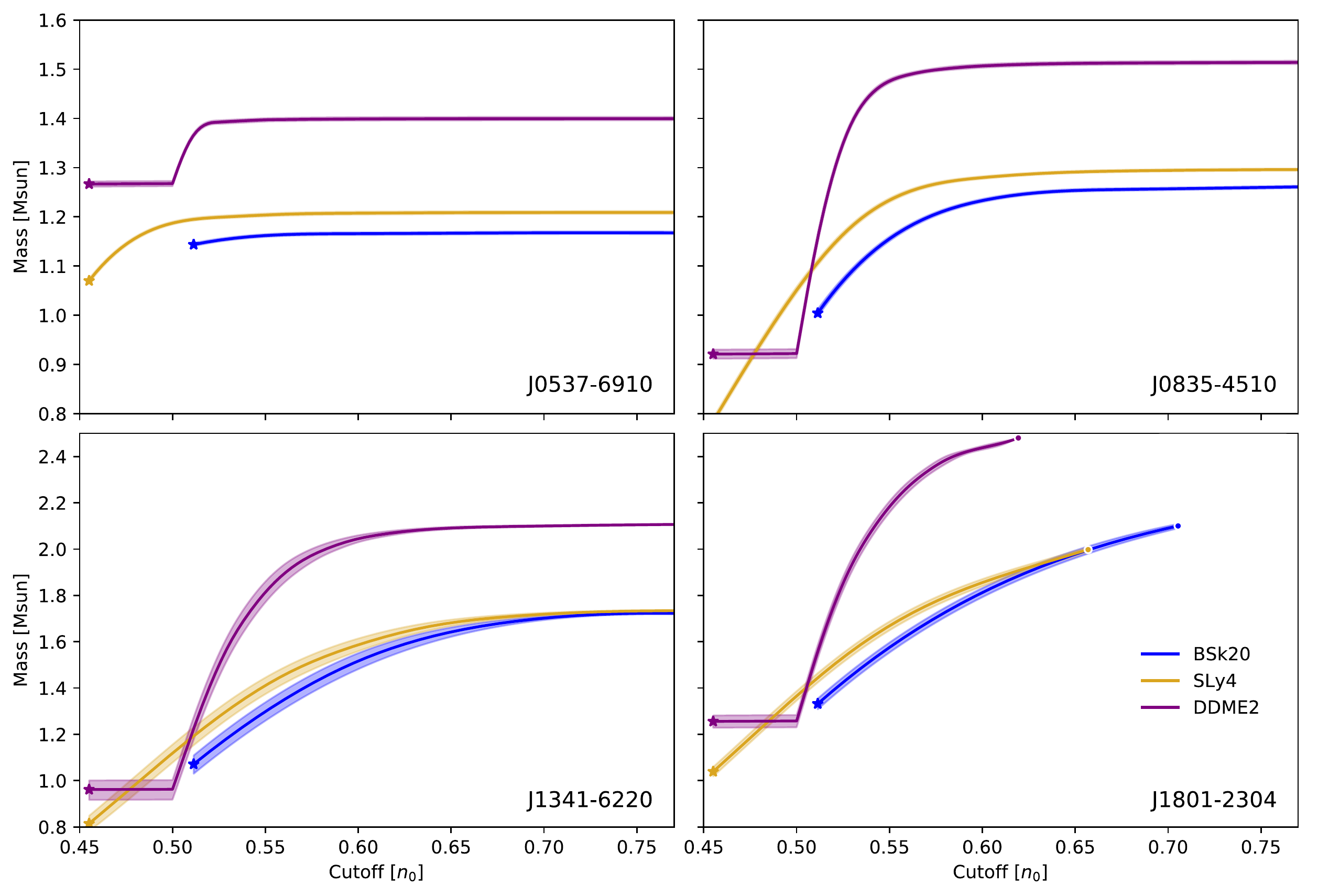}
\caption{
The mass $M_{\rm act}$  as a function of the cutoff	baryon density for the superfluid reservoir, $n_{\rm cut}$, for the four pulsars in the sample which have $N_{\rm m}>4$ and for the three EoSs considered: SLy4 (yellow), BSk20 (blue) and DDME2 (purple). 
All lines start with a pulsar, which indicates the cutoff $n_{\rm cut}=n_c$ at the crust-core interface (crust-only case). In the case of J1801-2304, the lines end with a dot at $M_{\rm eos}$, the maximum mass allowed by each EoS. The shaded regions indicate the uncertainty on the mass estimate. The plateau in the DDME2 curves is a by-product of the presence of a strong first-order phase transition at the core-crust interface.}
\label{fig:quadratelli}
\end{figure*}


To better follow the dependence of the mass estimates on the cutoff, in Fig~\ref{fig:quadratelli} we consider  the four benchmark pulsars  with $N_{\rm m} > 4$. For each of these objects, we plot $M_{\rm act}$ as a function of the superfluid region cutoff $n_{\rm cut}$ for the three different EoSs considered.
The region of constant mass for DDME2   corresponds to the first-order phase transition mentioned before but the general trend of lower masses for smaller superfluid reservoir is preserved. Also, as expected, a stiffer EoS like DDME2 predicts larger masses than the two softer EoSs. 
Moreover, the star with the largest $\omega_{\rm act}^*$ in this figure, J0537-6910, shows small variability in mass between the cutoffs, reaching its plateau very soon (for $n_{\rm cut} > 0.55 n_0$), as opposed to the star with the smallest nominal lag, J1801-2304, which reaches the maximum mass allowed by each EoS well before $0.75\, n_0$ and shows no plateau. In these cases larger cutoffs just yield too much available reservoir of angular momentum, so that the pulsar is not constrained anymore: any mass $M_{\rm act}$ compatible with the EoS can produce its small observed glitches.


\section{Conclusions}
\label{sec:conclusions}


We studied the dependence of the upper bound to the mass of a glitching pulsar on the extension of the superfluid reservoir by considering the neutron superfluid involved in the glitch mechanism to be limited in spherical shells starting from neutron drip density and ending at different cutoff densities near the crust-core interface. 
The rationale behind this choice is that thermal effects may shrink the region where the superfluid resides (e.g. if a layer of normal matter exists between the singlet and the triplet neutron superfluid), thus reducing the associated angular momentum reservoir \citep{ho+2015}. However, the effects of temperature, which can occur as a variation of the extension of the superfluidity region or an enhancement of vortex creep inside the star, are neglected: differently from \citep{ho+2015}, we have assumed the same cutoff density for all the pulsars in the sample, irrespective of the different ages of the stars. Temperature effects could be included in the present framework by considering a more refined modelling of the coupled rotational and thermal evolution of a pulsar. 

We have chosen values for the cutoffs similar to those expected for the ${}^1S_0$ pairing gap, covering a range that takes into account the uncertainties of the theoretical calculations of pairing gaps and of the actual temperatures inside a glitching pulsar. As an extreme case, we have also considered a reservoir limited to the crust. Finally, we have compared the results to the case of superfluid extended to the whole star. By looking at Fig~\ref{fig:quadratelli}, it seems that to set more reliable constraint on the mass of a pulsar, it will be fundamental to understand the internal physics of neutron stars at densities between $0.5 \,  n_0$ and $0.6 \,  n_0$ (as this is the region at which the mass estimates are very sensitive to the cutoff of the superfluid region).

The present results are biased by some rough simplifications, that can however be relaxed in more refined studies, as discussed in Sec \ref{sec:model}. 
Here, we do not account for general relativistic effects, we consider only crustal pinning and we employ a unified toy-model for the dynamics of the lag between the two components. 
Nonetheless, our aim is to study the dependence of the mass upper bounds on the extension of the superfluid reservoir for a rather large sample pulsars, which have been selected for their statistical significance according to the parameter $N_{\rm m}$. 
The analysis of the sample indicates the four pulsars in Fig~\ref{fig:quadratelli} and other objects having $N_{\rm m} \gtrsim 3$, in particular J1740-3015 \citep{McKenna1990Nat}, are likely to provide the best opportunity to test the current understanding of the glitch phenomenon. 

The study of the pulsar sample also revealed a clear difference in the behaviour for different pulsars: in particular, glitchers with small $\omega_{\rm act}^*$ turn out to be strongly dependent on the extension of the superfluid reservoir. On the contrary, the masses of stars with large $\omega_{\rm act}^*$, such as the Crab (which is, however, exceptional with respect to the rest of the sample), do not depend on the cutoff considered. 

The present analysis confirms that the superfluid in the crust alone is not sufficient to explain the glitching activity of pulsars \citep{andersson+2012,chamel2013, delsate2016}. 
Altogether, reasonable values of $M_{\rm act}$, lying within the observational range measured for neutron star masses, are obtained only if the superfluid reservoir extends for at least a very small region inside the outer core, as also noticed by \cite{ho+2015} and \citep{gercinog2014}: this is compatible with several theoretical calculations of the neutron pairing gap.  
Moreover, the mass estimates stabilise when the reservoir reaches densities above nuclear saturation, indicating that there may be no need to consider parts of the core where the paring is expected to be in the P-wave.

We conclude by stressing that the exact values predicted for the masses are of no particular importance in the present context: changing the microscopic input (like the EoS, the pinning force and the entrainment parameter) can modify them to either bigger or smaller values, yielding a large degeneracy in the mass determination. 
What is conserved is just the  inverse correlation between amplitude of the largest observed glitch and the corresponding mass (which is analogous to the inverse correlation between the activity and the corresponding mass, see e.g. \citealt{LE99}): the  ``slope'' of the relation between $M_{\rm act}$ and $\Delta \Omega$ depends on the extension of the reservoir, while its ``height'' is determined by the microscopic input. 
In other words, the curves drawn by the points in Fig~\ref{fig:masses} must be calibrated. This may be possible in the future, if the masses of a few large glitchers are determined by some direct observation (e.g. with the discovery of a prolific glitcher in a binary system). In such a case, extended observations of pulsar glitches would allow to measure the mass of isolated neutron stars and, at the same time, would help to put some constraints on the microphysics of dense hadronic matter.

\section*{Acknowledgements}

Partial support comes from The Multi-messenger Physics and
Astrophysics of Neutron Stars (PHAROS), COST Action CA16214.
MA acknowledges support from the Narodowe Centrum Nauki
(NCN) grant SONATA BIS 2015/18/E/ST9/00577, P.I.: B. Haskell.
We thank Nils Andersson for helpful suggestions and Morgane
Fortin for providing the SLy4+DDME2 equation of state.

\bibliographystyle{mnras}
\bibliography{biblio}

\begin{thebibliography}{}
\makeatletter
\relax
\def\mn@urlcharsother{\let\do\@makeother \do\$\do\&\do\#\do\^\do\_\do\%\do\~}
\def\mn@doi{\begingroup\mn@urlcharsother \@ifnextchar [ {\mn@doi@}
  {\mn@doi@[]}}
\def\mn@doi@[#1]#2{\def\@tempa{#1}\ifx\@tempa\@empty \href
  {http://dx.doi.org/#2} {doi:#2}\else \href {http://dx.doi.org/#2} {#1}\fi
  \endgroup}
\def\mn@eprint#1#2{\mn@eprint@#1:#2::\@nil}
\def\mn@eprint@arXiv#1{\href {http://arxiv.org/abs/#1} {{\tt arXiv:#1}}}
\def\mn@eprint@dblp#1{\href {http://dblp.uni-trier.de/rec/bibtex/#1.xml}
  {dblp:#1}}
\def\mn@eprint@#1:#2:#3:#4\@nil{\def\@tempa {#1}\def\@tempb {#2}\def\@tempc
  {#3}\ifx \@tempc \@empty \let \@tempc \@tempb \let \@tempb \@tempa \fi \ifx
  \@tempb \@empty \def\@tempb {arXiv}\fi \@ifundefined
  {mn@eprint@\@tempb}{\@tempb:\@tempc}{\expandafter \expandafter \csname
  mn@eprint@\@tempb\endcsname \expandafter{\@tempc}}}

\bibitem[\protect\citeauthoryear{{Alpar}, {Langer}  \& {Sauls}}{{Alpar}
  et~al.}{1984}]{alpar84rapid}
{Alpar} M.~A.,  {Langer} S.~A.,   {Sauls} J.~A.,  1984, \mn@doi [ApJ]
  {10.1086/162232}, \href {http://adsabs.harvard.edu/abs/1984ApJ...282..533A}
  {282, 533}

\bibitem[\protect\citeauthoryear{{Anderson} \& {Itoh}}{{Anderson} \&
  {Itoh}}{1975}]{andersonitoh1975}
{Anderson} P.~W.,  {Itoh} N.,  1975, \mn@doi [\nat] {10.1038/256025a0}, \href
  {http://adsabs.harvard.edu/abs/1975Natur.256...25A} {256, 25}

\bibitem[\protect\citeauthoryear{{Andersson} \& {Comer}}{{Andersson} \&
  {Comer}}{2006}]{andersson2006CQGra}
{Andersson} N.,  {Comer} G.~L.,  2006, \mn@doi [Classical and Quantum Gravity]
  {10.1088/0264-9381/23/18/003}, \href
  {https://ui.adsabs.harvard.edu/abs/2006CQGra..23.5505A} {23, 5505}

\bibitem[\protect\citeauthoryear{{Andersson}, {Sidery}  \& {Comer}}{{Andersson}
  et~al.}{2006}]{AndSid06}
{Andersson} N.,  {Sidery} T.,   {Comer} G.~L.,  2006, \mn@doi [\mnras]
  {10.1111/j.1365-2966.2006.10147.x}, \href
  {http://ukads.nottingham.ac.uk/abs/2006MNRAS.368..162A} {368, 162}

\bibitem[\protect\citeauthoryear{{Andersson}, {Glampedakis}, {Ho}  \&
  {Espinoza}}{{Andersson} et~al.}{2012}]{andersson+2012}
{Andersson} N.,  {Glampedakis} K.,  {Ho} W.~C.~G.,   {Espinoza} C.~M.,  2012,
  \mn@doi [Physical Review Letters] {10.1103/PhysRevLett.109.241103}, \href
  {http://adsabs.harvard.edu/abs/2012PhRvL.109x1103A} {109, 241103}

\bibitem[\protect\citeauthoryear{{Andreev} \& {Bashkin}}{{Andreev} \&
  {Bashkin}}{1975}]{andreevbashkin1975}
{Andreev} A.~F.,  {Bashkin} E.~P.,  1975, Soviet Physics - JETP, \href
  {http://adsabs.harvard.edu/abs/1976JETP...42..164A} {42, 164}

\bibitem[\protect\citeauthoryear{{Antonelli} \& {Pizzochero}}{{Antonelli} \&
  {Pizzochero}}{2017}]{antonellipizzochero2017}
{Antonelli} M.,  {Pizzochero} P.~M.,  2017, \mn@doi [\mnras]
  {10.1093/mnras/stw2376}, \href
  {http://adsabs.harvard.edu/abs/2017MNRAS.464..721A} {464, 721}

\bibitem[\protect\citeauthoryear{{Antonelli}, {Montoli}  \&
  {Pizzochero}}{{Antonelli} et~al.}{2018}]{antonelli+2018}
{Antonelli} M.,  {Montoli} A.,   {Pizzochero} P.~M.,  2018, \mn@doi [\mnras]
  {10.1093/mnras/sty130}, \href
  {http://adsabs.harvard.edu/abs/2018MNRAS.475.5403A} {475, 5403}

\bibitem[\protect\citeauthoryear{{Antonopoulou}, {Espinoza}, {Kuiper}  \&
  {Andersson}}{{Antonopoulou} et~al.}{2018}]{antonJ0537}
{Antonopoulou} D.,  {Espinoza} C.~M.,  {Kuiper} L.,   {Andersson} N.,  2018,
  \mn@doi [\mnras] {10.1093/mnras/stx2429}, \href
  {http://adsabs.harvard.edu/abs/2018MNRAS.473.1644A} {473, 1644}

\bibitem[\protect\citeauthoryear{{Ashton}, {Prix}  \& {Jones}}{{Ashton}
  et~al.}{2017}]{ashton2017}
{Ashton} G.,  {Prix} R.,   {Jones} D.~I.,  2017, \mn@doi [\prd]
  {10.1103/PhysRevD.96.063004}, \href
  {http://adsabs.harvard.edu/abs/2017PhRvD..96f3004A} {96, 063004}

\bibitem[\protect\citeauthoryear{{Ashton}, {Lasky}, {Graber}  \&
  {Palfreyman}}{{Ashton} et~al.}{2019}]{ashton2019NatAs}
{Ashton} G.,  {Lasky} P.~D.,  {Graber} V.,   {Palfreyman} J.,  2019, \mn@doi
  [Nature Astronomy] {10.1038/s41550-019-0844-6}, \href
  {https://ui.adsabs.harvard.edu/abs/2019NatAs...3.1143A} {3, 1143}

\bibitem[\protect\citeauthoryear{{Baym}, {Pethick}, {Pines}  \&
  {Ruderman}}{{Baym} et~al.}{1969}]{baym+1969}
{Baym} G.,  {Pethick} C.,  {Pines} D.,   {Ruderman} M.,  1969, \mn@doi [\nat]
  {10.1038/224872a0}, \href {http://adsabs.harvard.edu/abs/1969Natur.224..872B}
  {224, 872}

\bibitem[\protect\citeauthoryear{{Carter}, {Chamel}  \& {Haensel}}{{Carter}
  et~al.}{2005}]{carter2005NuPhA}
{Carter} B.,  {Chamel} N.,   {Haensel} P.,  2005, \mn@doi [\nphysa]
  {10.1016/j.nuclphysa.2004.11.006}, \href
  {https://ui.adsabs.harvard.edu/abs/2005NuPhA.748..675C} {748, 675}

\bibitem[\protect\citeauthoryear{{Carter}, {Chamel}  \& {Haensel}}{{Carter}
  et~al.}{2006}]{carter2006IJMPD}
{Carter} B.,  {Chamel} N.,   {Haensel} P.,  2006, \mn@doi [International
  Journal of Modern Physics D] {10.1142/S0218271806008504}, \href
  {https://ui.adsabs.harvard.edu/abs/2006IJMPD..15..777C} {15, 777}

\bibitem[\protect\citeauthoryear{{Chamel}}{{Chamel}}{2012}]{chamel2012}
{Chamel} N.,  2012, \mn@doi [\prc] {10.1103/PhysRevC.85.035801}, \href
  {http://adsabs.harvard.edu/abs/2012PhRvC..85c5801C} {85, 035801}

\bibitem[\protect\citeauthoryear{{Chamel}}{{Chamel}}{2013}]{chamel2013}
{Chamel} N.,  2013, \mn@doi [Physical Review Letters]
  {10.1103/PhysRevLett.110.011101}, \href
  {http://adsabs.harvard.edu/abs/2013PhRvL.110a1101C} {110, 011101}

\bibitem[\protect\citeauthoryear{{Chamel}}{{Chamel}}{2017}]{chamel_review2017JApA}
{Chamel} N.,  2017, \mn@doi [Journal of Astrophysics and Astronomy]
  {10.1007/s12036-017-9470-9}, \href
  {https://ui.adsabs.harvard.edu/abs/2017JApA...38...43C} {38, 43}

\bibitem[\protect\citeauthoryear{{Chamel} \& {Carter}}{{Chamel} \&
  {Carter}}{2006}]{chamelcarter2006}
{Chamel} N.,  {Carter} B.,  2006, \mn@doi [\mnras]
  {10.1111/j.1365-2966.2006.10170.x}, \href
  {https://ui.adsabs.harvard.edu/abs/2006MNRAS.368..796C} {368, 796}

\bibitem[\protect\citeauthoryear{{Chamel} \& {Haensel}}{{Chamel} \&
  {Haensel}}{2006}]{chamelhaensel2006}
{Chamel} N.,  {Haensel} P.,  2006, \mn@doi [\prc] {10.1103/PhysRevC.73.045802},
  \href {http://adsabs.harvard.edu/abs/2006PhRvC..73d5802C} {73, 045802}

\bibitem[\protect\citeauthoryear{{Chamel} \& {Haensel}}{{Chamel} \&
  {Haensel}}{2008}]{chamelReview}
{Chamel} N.,  {Haensel} P.,  2008, \mn@doi [Living Reviews in Relativity]
  {10.12942/lrr-2008-10}, \href
  {http://adsabs.harvard.edu/abs/2008LRR....11...10C} {11, 10}

\bibitem[\protect\citeauthoryear{{Datta} \& {Alpar}}{{Datta} \&
  {Alpar}}{1993}]{dattaalpar1993}
{Datta} B.,  {Alpar} M.~A.,  1993, \aap, \href
  {http://adsabs.harvard.edu/abs/1993A%26A...275..210D} {275, 210}

\bibitem[\protect\citeauthoryear{{Delsate}, {Chamel}, {G{\"u}rlebeck},
  {Fantina}, {Pearson}  \& {Ducoin}}{{Delsate} et~al.}{2016}]{delsate2016}
{Delsate} T.,  {Chamel} N.,  {G{\"u}rlebeck} N.,  {Fantina} A.~F.,  {Pearson}
  J.~M.,   {Ducoin} C.,  2016, \mn@doi [\prd] {10.1103/PhysRevD.94.023008},
  \href {http://adsabs.harvard.edu/abs/2016PhRvD..94b3008D} {94, 023008}

\bibitem[\protect\citeauthoryear{{Douchin} \& {Haensel}}{{Douchin} \&
  {Haensel}}{2001}]{douchinhaensel2001}
{Douchin} F.,  {Haensel} P.,  2001, \mn@doi [\aap]
  {10.1051/0004-6361:20011402}, \href
  {http://adsabs.harvard.edu/abs/2001A%26A...380..151D} {380, 151}

\bibitem[\protect\citeauthoryear{{Easson}}{{Easson}}{1979}]{Easson1979}
{Easson} I.,  1979, \mn@doi [\apj] {10.1086/156842}, \href
  {http://adsabs.harvard.edu/abs/1979ApJ...228..257E} {228, 257}

\bibitem[\protect\citeauthoryear{{Espinoza}, {Lyne}, {Stappers}  \&
  {Kramer}}{{Espinoza} et~al.}{2011}]{espinoza+2011}
{Espinoza} C.~M.,  {Lyne} A.~G.,  {Stappers} B.~W.,   {Kramer} M.,  2011,
  \mn@doi [\mnras] {10.1111/j.1365-2966.2011.18503.x}, \href
  {http://adsabs.harvard.edu/abs/2011MNRAS.414.1679E} {414, 1679}

\bibitem[\protect\citeauthoryear{{Fortin}, {Provid{\^e}ncia}, {Raduta},
  {Gulminelli}, {Zdunik}, {Haensel}  \& {Bejger}}{{Fortin}
  et~al.}{2016}]{fortinC16}
{Fortin} M.,  {Provid{\^e}ncia} C.,  {Raduta} A.~R.,  {Gulminelli} F.,
  {Zdunik} J.~L.,  {Haensel} P.,   {Bejger} M.,  2016, \mn@doi [\prc]
  {10.1103/PhysRevC.94.035804}, \href
  {http://esoads.eso.org/abs/2016PhRvC..94c5804F} {94, 035804}

\bibitem[\protect\citeauthoryear{{Fuentes}, {Espinoza}, {Reisenegger}, {Shaw},
  {Stappers}  \& {Lyne}}{{Fuentes} et~al.}{2017}]{fuentes17}
{Fuentes} J.~R.,  {Espinoza} C.~M.,  {Reisenegger} A.,  {Shaw} B.,  {Stappers}
  B.~W.,   {Lyne} A.~G.,  2017, \mn@doi [\aap] {10.1051/0004-6361/201731519},
  \href {https://ui.adsabs.harvard.edu/#abs/2017A&A...608A.131F} {608, A131}

\bibitem[\protect\citeauthoryear{{Goriely}, {Chamel}  \& {Pearson}}{{Goriely}
  et~al.}{2010}]{goriely+2010}
{Goriely} S.,  {Chamel} N.,   {Pearson} J.~M.,  2010, \mn@doi [\prc]
  {10.1103/PhysRevC.82.035804}, \href
  {http://adsabs.harvard.edu/abs/2010PhRvC..82c5804G} {82, 035804}

\bibitem[\protect\citeauthoryear{{Graber}, {Cumming}  \& {Andersson}}{{Graber}
  et~al.}{2018}]{graber+2018}
{Graber} V.,  {Cumming} A.,   {Andersson} N.,  2018, \mn@doi [\apj]
  {10.3847/1538-4357/aad776}, \href
  {https://ui.adsabs.harvard.edu/abs/2018ApJ...865...23G} {865, 23}

\bibitem[\protect\citeauthoryear{{G{\"u}gercino{\u{g}}lu} \&
  {Alpar}}{{G{\"u}gercino{\u{g}}lu} \& {Alpar}}{2014}]{gercinog2014}
{G{\"u}gercino{\u{g}}lu} E.,  {Alpar} M.~A.,  2014, arXiv e-prints, \href
  {https://ui.adsabs.harvard.edu/abs/2014arXiv1405.6635G} {p. arXiv:1405.6635}

\bibitem[\protect\citeauthoryear{{Hartle}}{{Hartle}}{1967}]{hartle67}
{Hartle} J.~B.,  1967, \mn@doi [\apj] {10.1086/149400}, \href
  {http://adsabs.harvard.edu/abs/1967ApJ...150.1005H} {150, 1005}

\bibitem[\protect\citeauthoryear{{Haskell} \& {Melatos}}{{Haskell} \&
  {Melatos}}{2015}]{haskellmelatos2015}
{Haskell} B.,  {Melatos} A.,  2015, \mn@doi [International Journal of Modern
  Physics D] {10.1142/S0218271815300086}, \href
  {http://adsabs.harvard.edu/abs/2015IJMPD..2430008H} {24, 1530008}

\bibitem[\protect\citeauthoryear{{Haskell} \& {Sedrakian}}{{Haskell} \&
  {Sedrakian}}{2017}]{haskellsedrakian2017}
{Haskell} B.,  {Sedrakian} A.,  2017, preprint, \href
  {http://adsabs.harvard.edu/abs/2017arXiv170910340H} {} (\mn@eprint {arXiv}
  {1709.10340})

\bibitem[\protect\citeauthoryear{{Haskell}, {Pizzochero}  \&
  {Sidery}}{{Haskell} et~al.}{2012}]{hasekll2012}
{Haskell} B.,  {Pizzochero} P.~M.,   {Sidery} T.,  2012, \mn@doi [\mnras]
  {10.1111/j.1365-2966.2011.20080.x}, \href
  {https://ui.adsabs.harvard.edu/abs/2012MNRAS.420..658H} {420, 658}

\bibitem[\protect\citeauthoryear{{Ho}, {Espinoza}, {Antonopoulou}  \&
  {Andersson}}{{Ho} et~al.}{2015}]{ho+2015}
{Ho} W.~C.~G.,  {Espinoza} C.~M.,  {Antonopoulou} D.,   {Andersson} N.,  2015,
  \mn@doi [Science Advances] {10.1126/sciadv.1500578}, \href
  {http://adsabs.harvard.edu/abs/2015SciA....1E0578H} {1, e1500578}

\bibitem[\protect\citeauthoryear{{Howitt}, {Melatos}  \& {Delaigle}}{{Howitt}
  et~al.}{2018}]{howitt2018ApJ}
{Howitt} G.,  {Melatos} A.,   {Delaigle} A.,  2018, \mn@doi [\apj]
  {10.3847/1538-4357/aae20a}, \href
  {https://ui.adsabs.harvard.edu/abs/2018ApJ...867...60H} {867, 60}

\bibitem[\protect\citeauthoryear{{Khalatnikov}}{{Khalatnikov}}{1957}]{Khalatnikov57}
{Khalatnikov} I.~M.,  1957, Soviet Physics - JETP, 5, 542

\bibitem[\protect\citeauthoryear{{Khomenko} \& {Haskell}}{{Khomenko} \&
  {Haskell}}{2018}]{khomenko2018}
{Khomenko} V.,  {Haskell} B.,  2018, \mn@doi [\pasa] {10.1017/pasa.2018.12},
  \href {http://adsabs.harvard.edu/abs/2018PASA...35...20K} {35, e020}

\bibitem[\protect\citeauthoryear{{Lalazissis}, {Nik{\v s}i{\'c}}, {Vretenar}
  \& {Ring}}{{Lalazissis} et~al.}{2005}]{LNV2005}
{Lalazissis} G.~A.,  {Nik{\v s}i{\'c}} T.,  {Vretenar} D.,   {Ring} P.,  2005,
  \mn@doi [\prc] {10.1103/PhysRevC.71.024312}, \href
  {http://esoads.eso.org/abs/2005PhRvC..71b4312L} {71, 024312}

\bibitem[\protect\citeauthoryear{{Link}, {Epstein}  \& {Lattimer}}{{Link}
  et~al.}{1999}]{LE99}
{Link} B.,  {Epstein} R.~I.,   {Lattimer} J.~M.,  1999, \mn@doi [Physical
  Review Letters] {10.1103/PhysRevLett.83.3362}, \href
  {http://adsabs.harvard.edu/abs/1999PhRvL..83.3362L} {83, 3362}

\bibitem[\protect\citeauthoryear{Lyne, Shemar  \& Graham~Smith}{Lyne
  et~al.}{2000}]{lyne2000}
Lyne A.~G.,  Shemar S.~L.,   Graham~Smith F.,  2000, \mn@doi [Monthly Notices
  of the Royal Astronomical Society] {10.1046/j.1365-8711.2000.03415.x}, 315,
  534

\bibitem[\protect\citeauthoryear{{Manchester}, {Hobbs}, {Teoh}  \&
  {Hobbs}}{{Manchester} et~al.}{2005}]{manchester+2005}
{Manchester} R.~N.,  {Hobbs} G.~B.,  {Teoh} A.,   {Hobbs} M.,  2005, \mn@doi
  [\aj] {10.1086/428488}, \href
  {http://adsabs.harvard.edu/abs/2005AJ....129.1993M} {129, 1993}

\bibitem[\protect\citeauthoryear{{McKenna} \& {Lyne}}{{McKenna} \&
  {Lyne}}{1990}]{McKenna1990Nat}
{McKenna} J.,  {Lyne} A.~G.,  1990, \mn@doi [\nat] {10.1038/343349a0}, \href
  {https://ui.adsabs.harvard.edu/abs/1990Natur.343..349M} {343, 349}

\bibitem[\protect\citeauthoryear{{Melatos}, {Peralta}  \& {Wyithe}}{{Melatos}
  et~al.}{2008}]{melatos2008}
{Melatos} A.,  {Peralta} C.,   {Wyithe} J.~S.~B.,  2008, \mn@doi [\apj]
  {10.1086/523349}, \href {http://adsabs.harvard.edu/abs/2008ApJ...672.1103M}
  {672, 1103}

\bibitem[\protect\citeauthoryear{{Melatos}, {Howitt}  \& {Fulgenzi}}{{Melatos}
  et~al.}{2018}]{MHF2018}
{Melatos} A.,  {Howitt} G.,   {Fulgenzi} W.,  2018, \mn@doi [\apj]
  {10.3847/1538-4357/aad228}, \href
  {http://adsabs.harvard.edu/abs/2018ApJ...863..196M} {863, 196}

\bibitem[\protect\citeauthoryear{{Palfreyman}, {Dickey}, {Hotan}, {Ellingsen}
  \& {van Straten}}{{Palfreyman} et~al.}{2018}]{palfreyman2018Nat}
{Palfreyman} J.,  {Dickey} J.~M.,  {Hotan} A.,  {Ellingsen} S.,   {van Straten}
  W.,  2018, \mn@doi [\nat] {10.1038/s41586-018-0001-x}, \href
  {https://ui.adsabs.harvard.edu/abs/2018Natur.556..219P} {556, 219}

\bibitem[\protect\citeauthoryear{{Pizzochero}}{{Pizzochero}}{2011}]{pizzochero2011}
{Pizzochero} P.~M.,  2011, \mn@doi [\apjl] {10.1088/2041-8205/743/1/L20}, \href
  {http://adsabs.harvard.edu/abs/2011ApJ...743L..20P} {743, L20}

\bibitem[\protect\citeauthoryear{{Pizzochero}, {Antonelli}, {Haskell}  \&
  {Seveso}}{{Pizzochero} et~al.}{2017}]{pizzochero+2017}
{Pizzochero} P.~M.,  {Antonelli} M.,  {Haskell} B.,   {Seveso} S.,  2017,
  \mn@doi [Nature Astronomy] {10.1038/s41550-017-0134}, \href
  {http://adsabs.harvard.edu/abs/2017NatAs...1E.134P} {1, 0134}

\bibitem[\protect\citeauthoryear{{Pizzochero}, {Montoli}  \&
  {Antonelli}}{{Pizzochero} et~al.}{2019}]{lettera_overshoot}
{Pizzochero} P.,  {Montoli} A.,   {Antonelli} M.,  2019, arXiv e-prints, \href
  {https://ui.adsabs.harvard.edu/abs/2019arXiv191000066P} {p. arXiv:1910.00066}

\bibitem[\protect\citeauthoryear{{Prix}}{{Prix}}{2004}]{prix2004}
{Prix} R.,  2004, \mn@doi [\prd] {10.1103/PhysRevD.69.043001}, \href
  {http://adsabs.harvard.edu/abs/2004PhRvD..69d3001P} {69, 043001}

\bibitem[\protect\citeauthoryear{{Prix}, {Comer}  \& {Andersson}}{{Prix}
  et~al.}{2002}]{prix2002}
{Prix} R.,  {Comer} G.~L.,   {Andersson} N.,  2002, \mn@doi [\aap]
  {10.1051/0004-6361:20011499}, \href
  {https://ui.adsabs.harvard.edu/abs/2002A&A...381..178P} {381, 178}

\bibitem[\protect\citeauthoryear{{Ruderman} \& {Sutherland}}{{Ruderman} \&
  {Sutherland}}{1974}]{ruderShut1974}
{Ruderman} M.~A.,  {Sutherland} P.~G.,  1974, \mn@doi [\apj] {10.1086/152857},
  \href {https://ui.adsabs.harvard.edu/abs/1974ApJ...190..137R} {190, 137}

\bibitem[\protect\citeauthoryear{{Schwenk}, {Friman}  \& {Brown}}{{Schwenk}
  et~al.}{2003}]{schwenk+2003}
{Schwenk} A.,  {Friman} B.,   {Brown} G.~E.,  2003, \mn@doi [Nuclear Physics A]
  {10.1016/S0375-9474(02)01290-3}, \href
  {http://adsabs.harvard.edu/abs/2003NuPhA.713..191S} {713, 191}

\bibitem[\protect\citeauthoryear{{Sedrakian} \& {Shakhabasian}}{{Sedrakian} \&
  {Shakhabasian}}{1980}]{Sedrakian1980}
{Sedrakian} D.~M.,  {Shakhabasian} K.~M.,  1980, Astrofizika, \href
  {https://ui.adsabs.harvard.edu/abs/1980Afz....16..727S} {16, 727}

\bibitem[\protect\citeauthoryear{{Seveso}, {Pizzochero}, {Grill}  \&
  {Haskell}}{{Seveso} et~al.}{2016}]{seveso+2016}
{Seveso} S.,  {Pizzochero} P.~M.,  {Grill} F.,   {Haskell} B.,  2016, \mn@doi
  [\mnras] {10.1093/mnras/stv2579}, \href
  {http://adsabs.harvard.edu/abs/2016MNRAS.455.3952S} {455, 3952}

\bibitem[\protect\citeauthoryear{{Shaw} et~al.,}{{Shaw}
  et~al.}{2018}]{shaw+2018}
{Shaw} B.,  et~al., 2018, \mn@doi [\mnras] {10.1093/mnras/sty1294}, \href
  {http://adsabs.harvard.edu/abs/2018MNRAS.478.3832S} {478, 3832}

\bibitem[\protect\citeauthoryear{{Sidery}, {Passamonti}  \&
  {Andersson}}{{Sidery} et~al.}{2010}]{sidery2010MNRA}
{Sidery} T.,  {Passamonti} A.,   {Andersson} N.,  2010, \mn@doi [\mnras]
  {10.1111/j.1365-2966.2010.16497.x}, \href
  {https://ui.adsabs.harvard.edu/abs/2010MNRAS.405.1061S} {405, 1061}

\bibitem[\protect\citeauthoryear{{Wong}, {Backer}  \& {Lyne}}{{Wong}
  et~al.}{2001}]{WongCrab}
{Wong} T.,  {Backer} D.~C.,   {Lyne} A.~G.,  2001, \mn@doi [\apj]
  {10.1086/318657}, \href {http://adsabs.harvard.edu/abs/2001ApJ...548..447W}
  {548, 447}

\bibitem[\protect\citeauthoryear{{Yu} et~al.,}{{Yu} et~al.}{2013}]{yu2013}
{Yu} M.,  et~al., 2013, \mn@doi [\mnras] {10.1093/mnras/sts366}, \href
  {http://adsabs.harvard.edu/abs/2013MNRAS.429..688Y} {429, 688}

\makeatother
\end{thebibliography}

\bsp
\label{lastpage}
\end{document}